\documentclass{aa}

\usepackage{graphicx, epstopdf, multicol, blindtext}
\usepackage{txfonts, color, caption, subcaption, paralist, longtable}
\usepackage[normalem]{ulem}
\usepackage{natbib}
\bibpunct{(}{)}{;}{a}{}{,}
\begin{document}

   \title{Helium ignition in rotating magnetized CO white dwarfs 
 leading to fast and faint rather than classical Type Ia supernovae}

   \author{P. Neunteufel
          \inst{1,2}
          \and
          S.-C. Yoon
          \inst{2}
          \and
          N. Langer
          \inst{1}
          }

   \institute{\inst{1}Argelander Institut f\"ur Astronomy (AIfA), University of Bonn,
              Auf dem H\"ugel 71, D-53121 Bonn\\
              \inst{2}Department of Physics and Astronomy, Seoul National University,
              599 Gwanak-ro, Gwanak-gu, Seoul, 151-742, Korea\\
              \email{neunteufel@astro.uni-bonn.de}}

   \date{Accepted for publication by A\&A}
\abstract
   {Based mostly on stellar models which do not include rotation,
CO white dwarfs which accrete helium at rates of about \mbox{$\sim 10^{-8}~\mathrm{M}_\odot/\mathrm{yr}$} have been 
put forward as candidate progenitors for a number of transient astrophysical phenomena, including 
supernovae of Type Ia, and the peculiar and fainter Type Iax supernovae.}
   {Here we study the impact of accretion-induced spin-up including the subsequent magnetic field generation,
angular momentum transport, and viscous heating on the white dwarf evolution up to the point of helium ignition.}
   {We resolve the structure of the helium accreting white dwarf models with a one dimensional Langrangian 
hydrodynamic code, modified to include rotational and magnetic effects, in 315 model sequences adopting different 
mass transfer rates ($10^{-8}\dots 10^{-7}~\mathrm{M}_\odot/\mathrm{yr}$), and initial white dwarf masses 
($0.54 \dots 1.10~\mathrm{M}_\odot$) and luminosities ($0.01\dots 1\,$L$_{\odot}$).}
   {We find magnetic angular momentum transport, which leads to quasi solid-body rotation, 
profoundly impacts the evolution of the white dwarf models, and the helium ignition conditions.
Our rotating lower mass ($0.54$ and $0.82~\mathrm{M}_\odot$) models accrete up to
50\% more mass up to ignition compared to the non-rotating case, while it is the opposite for our
more massive models. 
Furthermore, we find that rotation leads to up to 10-times smaller helium ignition densities,
except for the lowest adopted initial white dwarf mass.
Ignition densities of order $10^6\,$g/cm$^3$ are only found for the lowest accretion rates and
for large amounts of accreted helium ($\gtrsim 0.4\,\mathrm{M}_\odot$).
However, correspondingly massive donor stars would transfer mass at much higher rates.
We therefore expect explosive He-shell burning to mostly occur as deflagrations and
at $\dot{M}>2\cdot10^{-8}~\mathrm{M}_\odot/\mathrm{yr}$, regardless of white dwarf mass.}
   {Our results imply that helium accretion onto CO white dwarfs at the considered rates is unlikely to lead to 
the explosion of the CO core or to classical Type\,Ia supernovae,
but may instead produce events which belong to the recently identified classes of faint and fast hydrogen-free supernovae.}

   
   \titlerunning{Helium Ignitions on White Dwarfs}
   \maketitle
\section{Introduction}
For several decades, white dwarfs (WDs) have been considered a key component of the progenitors of a number of astrophysical transient phenomena. These include archetypical Type Ia supernovae (SNe Ia), Type Iax \citep[SNe Iax; see][]{LFA2003, FCC2013}, theoretically predicted subluminous SNe ".Ia" \citep{BSWN2007,KHG2014} and other "peculiar" transients \citep{WK2011}. Due to the key importance of SNe Ia in particular as a tool for high-redshift distance measurements \citep{PAG1999,RFC1998,SSP1998}, observationally disentangling these kinds of transients as well as theoretically understanding the differences between their production mechanisms would have potentially far reaching consequences.

Even in the light of numerous attempts, no direct progenitor of a SN Ia has so far been unambiguously identified, although a number of nondetections \citep[see][]{KFF2014,NVR2008,MM2008,SP2012,LBP2011} have considerably limited the available parameter space for candidate systems. Of those, \cite{LBP2011}, who attempted to identify the progenitor of the SN Ia 2011fe, is of particular interest with respect to the study of He-accreting models as it apparently excludes a large part of the available parameter space for He-donor stars but leaves open the possibility of a low-mass helium giant donor.

Interestingly the picture is somewhat clearer where SNe Iax are concerned. \cite{MJS2014} succeeded in identifying a possible progenitor of SN 2012Z as a luminous blue source, possibly a massive main sequence star or blue supergiant. However, a helium giant of $\sim2.0~\mathrm{M}_\odot$ is not excluded. The ambiguity of the source is manifested in the failed attempt by \cite{FDJ2015} to identify the progenitor of SN 2014dt, excluding some of the progenitor parameters found by \cite{MJS2014}, thereby hinting at the existence of a variety of SN Iax progenitors.

These findings point to a possible connection between helium accreting WDs as progenitors of SNe Ia and/or Iax. Observed cases have been modeled with some success as deflagrations of near-Chandrasekhar-mass WDs \citep{MKS2016,FJP2016,KOP2015}, while the observed properties of the companion stars imply high mass transfer rates ($\dot{M}\gtrsim10^{-6}~\mathrm{M}_\odot/\mathrm{yr}$). This leaves open the question whether systems with lower mass accretion rates \mbox{$\sim10^{-8}~\mathrm{M}_\odot/\mathrm{yr}$} and less-than near-Chandrasekhar-mass WDs could also be viable progenitors for any of these transients.

Recently, a possible connection between the theoretically predicted SNe .Ia and AM Canum-Venaticorum systems (AM CVns) \citep{W1995,N2005} was proposed \citep{SKW2010,BSWN2007,SB2009} and extensively studied \citep{SM2014}, suggesting a continuum of He-accretion-induced transients defined by the nature of the donor star, the system's orbital period and the associated mass transfer rate.

The idea that the accumulation of material at rates of \mbox{$\sim10^{-8}~\mathrm{M}_\odot/\mathrm{yr}$} from a helium rich companion on a sub-Chandrasekhar mass carbon-oxygen (CO) WD could lead to an explosive ignition was first seriously explored more than 30 years ago \citep{T1980,T1980b,N1980P,N1982a,N1982b}. While it was noted that some of these detonation left the CO core of the WD intact, producing a faint supernova, there was a possibility that the detonation of the He-envelope lead to a secondary detonation of the CO core, resulting in the complete disruption of the star (double detonation mechanism). Later one- and multidimensional studies \citep{L1990, LG1990, LG1991, LA1995, B1997P, L1997P, FHR2007, FRH2010, SRH2010, KSF2010, WK2011} established He-accretion as a promising avenue in the search for the progenitors of explosive transients like SNe Ia. However, synthetic spectra \citep{WK2011,WW1994,WTW1986,SRH2010,KSF2010} suggest that lower amounts of He generally lead to a better agreement with the spectra of regular SNe Ia. Studies trying to establish a link between naturally occurring binary systems and their ability to produce either SNe Iax or Ia using the double detonation mechanism were undertaken by \cite{WJH2013} and \cite{NYL2016}.

The state of the WD at ignition and the subsequent detonation or deflagration constitutes the crucial link between the progenitor and the resulting transient. The accretion behavior, including the amount of mass which a WD is able to steadily accrete before He ignition occurs, has previously been studied for the non-rotating case \citep[among others]{WK2011} and the rotating case \citep{YL2004a, YL2004b}. The latter two took into account the effects of a number of rotationally induced instabilities and the concurrent effects of angular momentum transport in differentially rotating stellar interiors. One of the most notable results of these studies was that, especially considering the effects of the dynamical shear instability, chemical mixing was strong and the amount of helium needed to induce ignition with an accretion rate of $\sim10^{-8}~\mathrm{M}_\odot/\mathrm{yr}$ was very low when compared to the non-rotating case because of rotational energy dissipation. This indicates that angular momentum dissipation significantly impacts the parameter space for possible progenitor systems.

One mechanism for angular momentum transport, not taken into account in the aforementioned studies of WD accretion, was proposed by \cite{S2002}. This mechanism (referred to as the Tayler-Spruit mechanism or dynamo in the following) relies on the Tayler instability \citep{T1973}, and occurs in the radiative parts of differentially rotating stars. Small seed fields may be amplified to a predominantly
torroidal magnetic field structure, with the associated magnetic torque leading to angular momentum transport
which reduces the shear.

The effects of this mechanism on stellar evolution has been studied for massive stars by \cite{MM2003} and \cite{HWS2005}. The case for low mass stars was investigated by \cite{MM2004}, and for the Sun by \cite{EMM2005}. Apart from having major implications for the evolution of non-degenerate stars, as discussed in these studies, it is concluded that the Tayler-Spruit mechanism tends to 
induce flat rotational profiles in the radiative zones of stars. It is therefore expected that the Tayler-Spruit mechanism 
could have a significant impact on the evolution of WDs, which, by their nature, are mostly non-convective.
Angular momentum transport as provided by the Tayler-Spruit mechanism appears indeed to be required to
understand the slow rotation of isolated neutron stars \cite{HWS2005} and white dwarfs \citep{SLP2008}, and the
spin-down of red giant cores as implied by asteroseismic observations \citep{MGB2012}.

This present study attempts to further refine our understanding of binary systems in which an accreting WD ignites its 
He-shell. To do this, we investigate the response of rotating CO WD models to helium accretion, 
taking into account the effects of the Tayler-Spruit mechanism. Each model is calculated up to the point of 
ignition of the accumulated helium envelope. We then discuss the impact of variations of initial model parameters on the mass of the helium envelope and densities at ignition and the implications of our results for observations.
 
This paper is organized as follows: In Sec.~\ref{sec:methods} we describe the utilized computational framework, the mathematical description of the physics under consideration ({Sec.~\ref{sec:numerical-models}) and the chosen initial models and input parameters (Sec.~\ref{sec:initial-models}). In Sec.~\ref{sec:results} we discuss the results obtained through our computations, first with an emphasis on the structural response of the WD to accretion (Sec.~\ref{sec:rotational-profile}), then with regards to the observational implications of its state at the point of helium ignition (Sec.~\ref{sec:ignitions}). 
We compare rotating models with and without including the Tayler-Spruit mechanism in Sect.\,4,
and discuss the fates of our simulated model WDs in Sect.\,5.
In Sec.~\ref{sec:conclusions} we discuss our findings, putting them into context with previously published works on the subject.

\section{Numerical methods and physical assumptions} \label{sec:methods}
\subsection{Numerical methods} \label{sec:numerical-models}
The Binary Evolution Code (BEC) is a well established computational framework capable of performing detailed one-dimensional numerical experiments of single or binary star systems \citep{LDWH2000, YL2004a}. The framework is capable of performing detailed hydrodynamical simulations of rotating and degenerate systems, including mass accretion \citep{HLW2000,HL2000}.
\subsubsection{Implementation of the Tayler-Spruit mechanism} \label{sec:tayler-spruit}
Angular momentum transport within a stellar model is assumed to follow a diffusive process \citep[as in][]{HLW2000, HL2000, ES1978,YL2005}, defined by a diffusion equation of the form
\begin{equation} \label{eq:diffusion-equation}
\left( \frac{\partial \omega}{\partial t} \right)_{M_r} = \frac{1}{i} \left( \frac{\partial}{\partial M_r} \right)_t \left[ \left( 4 \pi r^2 \rho)^2 \nu_{\mathrm{turb}} \right) \left( \frac{\partial \omega}{\partial M_r} \right)_t \right]~,
\end{equation}
where $\omega$ is the local angular velocity, $t$ time, $M_r$ the mass coordinate, $i$ the local specific moment of inertia, $r$ the radius, $\rho$ the local density and $\nu_\mathrm{turb}$ the local shear viscosity as defined by the active diffusive processes.

It is possible to accurately compute models of diffusive processes in non-degenerate stars using an explicit initial value approach \citep[see][]{HLW2000}. However, the sensitive dependence of some rotationally induced instabilities on the exact form of the angular velocity profile tends to lead to significant numerical errors in explicit schemes under certain circumstances. The pertinent problem here is encountered in cases of fast angular momentum redistribution occurring on a dynamical time scale, as encountered in studies of the dynamical shear instability. In order to deal with this problem, a nonlinear solving routine was implemented by \cite{YL2004a} in order to study the effects of dynamical shears. Further relevant mixing processes included in the framework are the secular shear instability, Goldreich-Schubert-Fricke instability and Eddington-Sweet circulation.}

Besides the dynamical shear instability, the Tayler-Spruit mechanism is capable of acting on dynamical timescales during the secular evolution of an accreting white dwarf. While an explicit treatment of the Tayler-Spruit mechanism, using the prescription by \cite{S2002}, was successfully used to study angular momentum transport in massive stars \citep{HWS2005}, the aforementioned problems regarding numerical stability preclude its use for the purposes of this study. Therefore, a new implicit numerical treatment is required, which was constructed on the basis of the implicit routines for the dynamical shear instability created by \cite{YL2004a}. We use essentially the same mathematical prescription as \cite{YL2004a}.

\cite{S2002} does not explicitly limit his derivations to non-degenerate matter, which is dealt with in this study through the use of appropriate material descriptions, as explained later in this section. However, the assumption of shellular rotation might not be entirely applicable to white dwarfs \citep[as discussed by][]{HLW2000,YL2004a}, although the angular momentum distribution of shellular rotation exhibits enough similarities with cylindrical rotation to serve as an adequate approximation.

It is assumed that the ordering of characteristic frequencies is such that
\begin{equation}
\omega_A \ll \omega \ll N,
\end{equation}
where $\omega_A$ is the well known Alfv\'{e}n frequency, $\omega$ is the angular velocity and $N$ is the Brunt-V\"{a}is\"{a}l\"{a}- (buoyancy-) frequency. The Brunt-V\"{a}is\"{a}l\"{a}-frequency can be decomposed into a chemical buoyancy frequency ($N_{\mu}$) and a thermal buoyancy frequency ($N_T$) such that
\begin{equation}
N^2=N_\mu^2+N_T^2~.
\end{equation}
The prescription assumes a locally stably stratified stellar interior, which translates into the condition that $N^2 > 0$.

In order for the dynamo to work, the local angular velocity gradient has to be large enough to overcome the stabilizing effects of the buoyancy forces. This is accomplished by demanding that the dimensionless differential rotation rate $q=\vert\frac{\partial \ln \omega}{\partial \ln r} \vert$ satisfies 
\begin{equation}
q>q_\mathrm{min}
\end{equation}
with
\begin{equation} \label{eq:minq}
q_{min} = q_0 + q_1
\end{equation} the minimal rotation rate for dynamo action.
$q_0$ and $q_1$ are the minimal rotation rates for the stabilizing effects of the chemical and thermal stratification respectively and can be written
\begin{equation} \label{eq:compn}
q_0 = \left( \frac{N_\mu}{\omega} \right)^{\frac{7}{4}} \left( \frac{\eta}{r^2 N_\mu} \right)^{\frac{1}{4}}
\end{equation}
and
\begin{equation}
q_1 = \left( \frac{N_T}{\omega} \right)^{\frac{7}{4}} \left( \frac{\eta}{r^2 N_T} \right)^{\frac{1}{4}} \left( \frac{\eta}{\kappa} \right)^{\frac{3}{4}}~ ,
\end{equation}
where $\eta$ is the magnetic diffusivity and $\kappa$ the thermal diffusivity.

We compute the magnetic diffusivity using the electrical conductivity according to \cite{NP1984}, which has been found to adequately describe the conditions in the non-degenerate as well as the degenerate parts of the star. The thermal diffusivity is computed taking into account the conditions in degenerate matter \citep[see][]{YL2004a}.

The "effective viscosity"\footnote{Note that, according to Eq.~\ref{eq:nue0} and Eq.~\ref{eq:nue1}, the magnitude of the turbulent viscosity may be independent of the shear, hence the term "effective viscosity".} for a predominantly chemically stable medium is given as 
\begin{equation} \label{eq:nue0}
\nu_{e0} = r^2 \omega q^2 \left( \frac{\omega}{N_\mu} \right)^4
\end{equation}
and for a predominantly thermally stable medium as
\begin{equation} \label{eq:nue1}
\nu_{e1} = r^2 \omega ~ \mathrm{max} \left[ \left( \frac{\omega}{N_T} \right)^{\frac{1}{2}} \left(\frac{\kappa}{r^2 N_T} \right)^\frac{1}{2} ~,~ q^2 \left( \frac{\omega}{N_T} \right)^4 \right] ~ .
\end{equation}
These expressions are then connected with a patching formula in the form
\begin{equation} \label{eq:viscosity}
\nu=\frac{\nu_{e0} \nu_{e1}}{\nu_{e0} + \nu_{e1}} f(q) ~, 
\end{equation}
where $\nu$ is assumed to represent the effective viscosity for a thermally and chemically stable medium, $f(q)$ is a term introduced to ensure that $\nu$ vanishes smoothly as $q$ approaches $q_\mathrm{min}$ and which is defined as
\begin{equation}
f(q)=
\begin{cases}
1-\frac{q_\mathrm{min}}{q} \quad &\mathrm{if} \quad q>q_\mathrm{min} \\
0 \quad &\mathrm{if} \quad q<q_\mathrm{min}
\end{cases} ~.
\end{equation}
Note that as $N_T$ and $N_\mu$ approach zero independently, indicating the medium becoming less stably stratified, $\nu_{e0}$ and $\nu_{e1}$ approach infinity. This means that, respectively, the other parameter will dominate Eq.~\ref{eq:viscosity}.
The thus computed effective viscosity is then summed up with those resulting from all other active diffusive processes in $\nu_\mathrm{turb}$ (see Eq.~(\ref{eq:diffusion-equation})).
Since the given Tayler-Spruit viscosity is singular in a number of variables, we introduce an artificial upper limit, corresponding to the local maximum of the shear viscosity attainable through the dynamical shear instability. The condition reads thus
\begin{equation}
\nu_\mathrm{max}=\frac{h_\mathrm{p}^2}{\tau_\mathrm{dyn}}~,
\end{equation} 
where $h_\mathrm{p}$ is the local pressure scale height and $\tau_\mathrm{dyn}$ the local dynamical timescale.
Since the shortest timescale any diffusive process in a star can act on is the dynamical timescale and the dynamical shear instability acts on this timescale it is a justifiable lower limit for the operating timescale of the Tayler-Spruit mechanism.

The rate of energy dissipation is estimated following \cite{KT1978} and \cite{ML1989} as
\begin{equation} \label{eq:ediss}
\epsilon_\mathrm{diss} = \frac{1}{2} \nu_\mathrm{turb}\sigma^2~,
\end{equation}
where $\sigma=\frac{\partial \omega}{\omega ~ \partial r}$ is the differential rotation rate (note: $q = \vert r \cdot \sigma \vert$). The value of $\nu_\mathrm{turb}$ changes on short timescales. For the purposes of the nonlinear diffusion solver, $\epsilon_\mathrm{diss}$ is calculated as the time average of the energy dissipation during each evolutionary timestep of the entire stellar evolution code.

It should be mentioned at this point that tests on massive star models produced good agreement between the upgraded implicit diffusion solver and the explicit solver used by \cite{HWS2005}.
\subsubsection{Equation of momentum balance} \label{sec:momentum-balance}
For later reference, we note that BEC implements centrifugal effects via the description provided by \cite{ES1978}. The equation of momentum balance is given as 
\begin{equation} \label{eq:mom-balance}
\left(\frac{\partial P}{\partial m_p} \right)_t = - \frac{G r_p^4}{4 \pi r_p^4} f_p - \frac{1}{4 \pi r_p^2} \left( \frac{\partial^2 r_p}{\partial t^2} \right)_{m_p} ,
\end{equation}
where $P$ is pressure, $t$ is time $r_p$ radius, $m_p$ mass, the latter two as contained within an equipotential surface. The effects of rotation are described by the quantity $f_p$, where
\begin{equation}
f_p = \frac{4 \pi m_p^4}{G m_p S_p}\langle g^{-1}\rangle^{-1}  ,
\end{equation}
with $S_p$ the area of the closed isobaric surface, $g$ the absolute of the local effective gravitational acceleration and $\langle g^{-1}\rangle$ the mean of its inverse. The correcting factor $f_p$ is obtained iteratively and consistent with the stellar structure equations \citep[see][]{ES1976}, but limited such that, locally, $f_p$ does not decrease anymore if $v/v_\mathrm{K} > 0.6$. This causes our methodology to generally underestimate effects relating to centrifugal forces wherever the local rotational velocity $v$ approaches the Keplerian value $v_\mathrm{K}$, which is defined as 
\begin{equation} \label{eq:kepler}
v_\mathrm{K}=\left(\frac{GM}{r}\right)^{1/2},
\end{equation}
where all variables are defined in the customary way.
\subsection{Initial models and input parameters} \label{sec:initial-models}
\begin{table}
\centering
\caption{Physical parameters of the initial white dwarf models:\\
Initial mass $M_\mathrm{WD,i}$, initial surface luminosity $L_\mathrm{s,i}$, initial core temperature $T_\mathrm{c,i}$, initial core density $\rho_\mathrm{c,i}$, initial radius $R_\mathrm{WD,i}$ and initial surface velocity $v_\mathrm{rot,i}$}
\label{tab:inmodel}
\begin{tabular}{l l l l l l}
\hline\hline
$M_\mathrm{WD,i}$ & $\log\left(L_\mathrm{s,i} / \mathrm{L}_\odot \right)$ & $T_\mathrm{c,i}$ & $\rho_\mathrm{c,i}$ & $R_\mathrm{WD,i}$ & $v_\mathrm{rot,i}$ \\
$M_\odot$ & & $10^7~\mathrm{K}$ & $10^6~\mathrm{g/cm}^3$ & $R_\odot$ & km/s \\
\hline 
0.54 & -0.006 & 0.621 & 3.33 & 0.0141 & 10.0 \\
0.54 & -0.998 & 0.427 & 3.52 & 0.0133 & 10.3 \\
0.54 & -2.003 & 0.176 & 3.65 & 0.0126 & 10.6 \\
\hline
0.82 & -0.004 & 0.660 & 11.4 & 0.0105 & 17.1 \\
0.82 & -1.024 & 0.417 & 11.8 & 0.0101 & 9.5 \\
0.82 & -1.998 & 0.168 & 12.1 & 0.0098 & 9.4 \\
\hline
1.00 & -0.001 & 0.783 & 31.7 & 0.0083 & 9.4 \\
1.00 & -1.000 & 0.414 & 32.8 & 0.0080 & 9.5 \\
1.00 & -1.995 & 0.155 & 33.6 & 0.0078 & 9.3 \\
\hline
1.10 & -0.011 & 0.858 & 65.5 & 0.0070 & 7.4 \\
1.10 & -1.017 & 0.317 & 68.5 & 0.0068 & 7.3 \\
\hline
\end{tabular}
\end{table}
In this study, we use CO WD models of masses $0.54~\mathrm{M}_\odot$, $0.82~\mathrm{M}_\odot$, $1.0~\mathrm{M}_\odot$ and $1.1~\mathrm{M}_\odot$. These were created by evolving a helium star of appropriate mass up to the end of the helium main-sequence and then artificially removing the majority of the remaining helium envelope. This results in a WD of about 60\% carbon and 40\% oxygen (though retaining a small, $<0.01~\mathrm{M}_\odot$ helium layer in case of the $0.82~\mathrm{M}_\odot$ and $1.0~\mathrm{M}_\odot$ models), an effective temperature of about $2 \cdot 10^5~\mathrm{K}$ and a surface luminosity of $\log\left( L/\mathrm{L}_\odot \right)\approx2.5$. The models are then spun up to a surface velocity of between $7~\mathrm{km/s}$ and $20~\mathrm{km/s}$, which is in agreement with measurements of the rotational velocities of isolated white dwarfs \citep[see][]{HNR1997, KDW1998, K2003, SLP2008, MGB2012}, and then allowed to cool to surface luminosities of about $\log\left( L/\mathrm{L}_\odot \right)=0~,~-1,~-2$ (see Table~\ref{tab:inmodel}). It is not immediately obvious whether these values are adequate. Since any He star+CO WD system will be a product of at least one (possibly two) common envelope event, the thermal history of the WD will be different from those usually obtained for isolated WDs \citep[e.g.][]{RAB2010}. However, since the details of common envelope evolution are currently far from being understood completely \citep[see][]{IJC2013}, the exact disposition of a WD at the onset of mass accretion is subject to some uncertainty. Enlarging the parameter space to include most likely initial luminosities is one way to address this problem. 
We note that the discrepancies in initial rotational velocities are expected to have negligible impact on the final result. Since the accreted material efficiently spins up the WD, surface velocities will reach values up to three magnitudes higher than the initial one long before any helium burning is expected to set in. For the purposes of this study, rotational velocities of the magnitude presented in Table~\ref{tab:inmodel} can be thought of as negligible. This will be clarified in greater detail in Sec.~\ref{sec:results}. Further, non-rotating models are computed for comparison.

WDs in binary systems are expected to accrete mass through a Keplerian accretion disc. The material falling onto the WD is therefore expected to move with a velocity close to the Keplerian value at the point of impact \citep{P1991,PN1991}, carrying a specific angular momentum of $j_\mathrm{kepler}$. Within the context of the BEC framework, the amount of specific angular momentum ($j_\mathrm{acc}$) carried by accreted matter is manually controlled by the choice of $f_\mathrm{acc}$ in the expression $j_\mathrm{acc}=f_\mathrm{acc}\cdot j_\mathrm{kepler}$. The choice of $f_\mathrm{acc}=1.0$ is the most natural one. In order to investigate conditions deviating from this assumption, we also include systems calculated with $f_\mathrm{acc}=0.6$ and $f_\mathrm{acc}=0.3$.

We avoid super-critical rotation by assuming that, if the star is rotating critically, no further angular momentum is accumulated, the excess momentum being dissipated in the accretion disk. This is accomplished by setting $f_\mathrm{acc}=0$ if further accretion would lead to super-critical rotation.

The accretion rates chosen in this study range between \mbox{$1\cdot10^{-8}~\mathrm{M}_\odot/\mathrm{yr}$} and $1\cdot10^{-7}~\mathrm{M}_\odot/\mathrm{yr}$, which are commensurate with average mass transfer rates achievable in a short period He star+CO WD system with the He star ($M\leq 1.0~\mathrm{M}_\odot$) undergoing He main sequence evolution. Consequently, the accreting material is assumed to consist of helium with an admixture of hydrogen-burning products corresponding to solar metallicity. At accretion rates higher than $1\cdot10^{-7}~\mathrm{M}_\odot/\mathrm{yr}$ the WD is expected to undergo a series of nova outbursts with individually comparatively low ejecta masses \citep{HK2004,PYT2015}. This mechanism is expected to have a profound impact on the outcome of the evolution of He star+CO WD systems \citep{NYL2016}, but both the nova cycle and the evolution of complete systems lie outside the intended scope of this work. 
For studies of the regime of accretion rates $\dot{M}\geq3 \cdot 10^{-6} \mathrm{M}_\odot/\mathrm{yr}$ the reader is referred to \cite{YL2003} and, more recently \cite{BBS2016}. In this regime, a period of steady burning and/or shell flashes potentially leads to a helium shell detonation, with the expectation of a detonation of the CO core of the WD.

\subsection{Ignition conditions} \label{sec:det-conditions}
All models are computed up to the point of unstable He ignition, which we define as the point where the helium burning timescale $\tau_\mathrm{nuc,He}$ approaches the dynamical timescale of the star (i.e.~$\tau_\mathrm{nuc,He}\approx \tau_\mathrm{dyn}$). Due to our methodology, we are unable to follow the ignition through to the final outcome of the process (i.e. detonation or deflagration). It is therefore necessary to apply some quantitative criterion in order to distinguish one outcome from the other. Possible criteria to distinguish detonations from deflagrations have been previously proposed by \cite{WW1994} and \cite{WK2011}. Both of these criteria use the density at the point of ignition as a measure as to whether the ignition would subsequently develop into a (supersonic) detonation or a (subsonic) deflagration. The defining factor in both cases is whether the density at the point of ignition is sufficient to preclude the premature quenching of the thermonuclear runaway through thermally induced expansion. The latter source introduces a critical density dependent on the ignition temperature. The former sets the critical density to a constant value of $\rho_\mathrm{crit}=1\cdot10^6~\mathrm{g/cm}^3$ while the latter gives 
\begin{equation}
\rho_{6,\mathrm{crit}}\gtrsim \left( \frac{0.0607}{T_8^2} \exp(\frac{20}{T_8}) \right)^{\frac{1}{2.3}}.
\end{equation}
However, as stated by \cite{WK2011}, the given function is an approximation which fails to properly identify detonating systems of the same study. The lowest predicted ignition density of all of the detonating systems in the aforementioned study is $\rho_\mathrm{ign}\simeq 6.8\cdot 10^{5} \mathrm{g}/\mathrm{cm}^3$ with most ignition densities exceeding $\rho_\mathrm{crit}=1\cdot10^6~\mathrm{g/cm}^3$. We therefore assume that the choice of a constant critical ignition density of $\rho_\mathrm{crit}=1\cdot10^6~\mathrm{g/cm}^3$ as put forward by \cite{WW1994} is justified for the purposes of this study.

\section{Results} \label{sec:results}
In this section, we will discuss the results obtained from our simulations. A selection of defining parameters of the systems under consideration (though limited to systems with $f_\mathrm{acc}=1.0$ for the sake of brevity) at the point of He-ignition are presented in Table~\ref{tab:results}. For the sake of clarity, it should be mentioned that in the following section, "rotating" always refers to models in the context of this study, i.e. rotating models with an active Tayler-Spruit dynamo, dynamical shear instability, Eddington-Sweet circulation and secular shear instability and "non-rotating" to non-rotating models with no angular momentum diffusion. Rotating models as defined here are explicitly compared with non-magnetic rotating models (i.e. rotating models without an active Tayler-Spruit dynamo) in Sec.~\ref{sec:comp}.
\begin{longtab} 
\begin{longtable}{l c c c c l l c l l l l}
\caption{This table lists the most important parameters of simulated systems with $f=1.0$ (see Sec.~\ref{sec:initial-models}) at the point of He-ignition. The leftmost column lists the model number. $L_\mathrm{init}$ is the luminosity at the start of mass accretion, $\dot{M}$ the accretion rate, $M_\mathrm{init}$ the initial mass of the model, $t$ the time since the start of mass accretion, $M_\mathrm{He,f}$ the mass of the accumulated He-envelope at the time of ignition, $T_\mathrm{4,f}$ the effective temperature at the time of ignition in units of $10^4~\mathrm{K}$, $L_\mathrm{f}$ the luminosity at ignition, $\rho_\mathrm{6,c}$ the core density at the time of ignition in units of $[10^6~\mathrm{g}/\mathrm{cm}^3]$, $T_\mathrm{8,ign}$ the temperature the the point of ignition in units of $[10^8~\mathrm{K}]$, $\rho_\mathrm{6,ign}$ the density at the point of ignition in units of $[10^6~\mathrm{g}/\mathrm{cm}^3]$ and $M_\mathrm{r,ign}$ the location of the point of ignition expressed in terms of the mass coordinate.
The model number is composed as follows: The first digit indicates the value of the initial luminosity, if preceded by the letter "m", that value should be understood to be negative. The second and third digit is equal to ten times value of the the initial mass rounded to the next $0.1~\mathrm{M}_\odot$. The fourth and fifth digit are equal to the value of ten times the chosen value for $f$. The hyphenated number is equal to the value of the chosen mass accretion rate divided by $10^{-8}$.}
\label{tab:results}\\
\hline
\hline
\# & $L_\mathrm{init}$ &	$\dot{M}/10^8$ &	$M_\mathrm{init}$&	$t$&	$M_\mathrm{He,f}$ &	$T_\mathrm{4,f}$	&	$L_\mathrm{f}$ &	$\rho_\mathrm{6,c}$	&	$T_\mathrm{8,ign}$	&	$\rho_\mathrm{6,ign}$	&	$M_\mathrm{r,ign}$	\\
 & $[\log(L/\mathrm{L}_\odot)]$	& $[\mathrm{M}_\odot/\mathrm{yr}]$	& $[\mathrm{M}_\odot]$	& $[\mathrm{Myr}]$	& $[\mathrm{M}_\odot]$	&		& $[\log(L/\mathrm{L}_\odot)]$	&	&	&	& $[\mathrm{M}_\odot]$	\\
\hline
\endfirsthead
\multicolumn{12}{l}%
{\tablename\ \thetable\ -- \textit{Continued from previous page}} \\
\hline
\hline
\# & $L_\mathrm{init}$ &	$\dot{M}$ &	$M_\mathrm{init}$&	$t$&	$M_\mathrm{He,f}$ &	$T_\mathrm{4,f}$	&	$L_\mathrm{f}$ &	$\rho_\mathrm{6,c}$	&	$T_\mathrm{8,ign}$	&	$\rho_\mathrm{6,ign}$	&	$M_\mathrm{r,ign}$	\\
\hline
\endhead
\hline \multicolumn{12}{l}{\textit{Continued on next page}} \\
\endfoot
\hline
\endlastfoot
Lm20510-1	&	-2	&	1	&	0.54	&	69.55	&	0.70	&	13.27	&	1.45	&	39.49	&	1.94	&	11.98	&	0.50	\\
Lm20510-2	&	-2	&	2	&	0.54	&	30.72	&	0.61	&	8.33	&	0.71	&	25.84	&	2.08	&	6.32	&	0.56	\\
Lm20510-3	&	-2	&	3	&	0.54	&	7.64	&	0.229	&	4.99	&	0.37	&	4.36	&	3.42	&	0.19	&	0.62	\\
Lm20510-4	&	-2	&	4	&	0.54	&	4.54	&	0.182	&	4.62	&	0.35	&	3.67	&	3.33	&	0.13	&	0.59	\\
Lm20510-5	&	-2	&	5	&	0.54	&	3.11	&	0.155	&	4.08	&	0.23	&	3.35	&	3.41	&	0.09	&	0.58	\\
Lm20510-6	&	-2	&	6	&	0.54	&	2.30	&	0.138	&	4.53	&	0.41	&	3.23	&	3.17	&	0.08	&	0.57	\\
Lm20510-7	&	-2	&	7	&	0.54	&	1.79	&	0.125	&	4.91	&	0.54	&	3.14	&	3.02	&	0.08	&	0.57	\\
Lm20510-8	&	-2	&	8	&	0.54	&	1.45	&	0.116	&	5.24	&	0.66	&	3.07	&	2.92	&	0.07	&	0.57	\\
Lm20510-9	&	-2	&	9	&	0.54	&	1.20	&	0.108	&	5.47	&	0.74	&	3.02	&	2.85	&	0.07	&	0.56	\\
Lm20510-10	&	-2	&	10	&	0.54	&	1.02	&	0.102	&	5.64	&	0.80	&	2.97	&	2.82	&	0.07	&	0.56	\\
\hline
Lm10510-1	&	-1	&	1	&	0.54	&	69.40	&	0.69	&	16.50	&	1.83	&	44.29	&	0.82	&	1.23	&	0.51	\\
Lm10510-2	&	-1	&	2	&	0.54	&	31.03	&	0.62	&	7.81	&	0.59	&	26.83	&	1.87	&	6.63	&	0.56	\\
Lm10510-3	&	-1	&	3	&	0.54	&	6.52	&	0.196	&	4.31	&	0.22	&	3.65	&	3.62	&	0.15	&	0.58	\\
Lm10510-4	&	-1	&	4	&	0.54	&	3.84	&	0.154	&	4.03	&	0.20	&	3.14	&	3.42	&	0.10	&	0.56	\\
Lm10510-5	&	-1	&	5	&	0.54	&	2.68	&	0.134	&	3.99	&	0.22	&	2.97	&	3.30	&	0.08	&	0.56	\\
Lm10510-6	&	-1	&	6	&	0.54	&	1.99	&	0.119	&	4.46	&	0.41	&	2.89	&	3.08	&	0.08	&	0.55	\\
Lm10510-7	&	-1	&	7	&	0.54	&	1.57	&	0.110	&	4.83	&	0.55	&	2.84	&	2.97	&	0.07	&	0.55	\\
Lm10510-8	&	-1	&	8	&	0.54	&	1.30	&	0.104	&	5.07	&	0.64	&	2.80	&	2.91	&	0.07	&	0.55	\\
Lm10510-9	&	-1	&	9	&	0.54	&	1.07	&	0.097	&	5.36	&	0.74	&	2.76	&	2.84	&	0.07	&	0.55	\\
Lm10510-10	&	-1	&	10	&	0.54	&	0.92	&	0.092	&	5.50	&	0.79	&	2.73	&	2.81	&	0.06	&	0.55	\\
\hline
L00510-1	&	0	&	1	&	0.54	&	69.50	&	0.69	&	12.69	&	1.37	&	39.27	&	1.88	&	11.98	&	0.49	\\
L00510-2	&	0	&	2	&	0.54	&	30.58	&	0.61	&	8.85	&	0.82	&	25.55	&	1.74	&	6.31	&	0.56	\\
L00510-3	&	0	&	3	&	0.54	&	5.58	&	0.167	&	4.78	&	0.36	&	3.34	&	3.01	&	0.17	&	0.56	\\
L00510-4	&	0	&	4	&	0.54	&	3.18	&	0.127	&	3.66	&	0.08	&	2.82	&	3.22	&	0.08	&	0.55	\\
L00510-5	&	0	&	5	&	0.54	&	2.17	&	0.109	&	4.24	&	0.33	&	2.69	&	2.95	&	0.07	&	0.55	\\
L00510-6	&	0	&	6	&	0.54	&	1.62	&	0.097	&	4.60	&	0.47	&	2.61	&	2.80	&	0.07	&	0.54	\\
L00510-7	&	0	&	7	&	0.54	&	1.25	&	0.088	&	4.91	&	0.59	&	2.55	&	2.68	&	0.06	&	0.54	\\
L00510-8	&	0	&	8	&	0.54	&	1.01	&	0.081	&	4.83	&	0.59	&	2.50	&	2.70	&	0.05	&	0.54	\\
L00510-9	&	0	&	9	&	0.54	&	0.87	&	0.078	&	4.98	&	0.67	&	2.49	&	2.72	&	0.05	&	0.54	\\
L00510-10	&	0	&	10	&	0.54	&	0.71	&	0.071	&	5.39	&	0.78	&	2.46	&	2.60	&	0.05	&	0.54	\\
\hline
\hline
Lm20810-1	&	-2	&	1	&	0.82	&	48.46	&	0.48	&	7.98	&	0.43	&	66.99	&	2.01	&	10.61	&	0.78	\\
Lm20810-2	&	-2	&	2	&	0.82	&	21.17	&	0.42	&	8.36	&	0.59	&	46.57	&	2.15	&	5.77	&	0.83	\\
Lm20810-3	&	-2	&	3	&	0.82	&	8.27	&	0.248	&	8.29	&	0.77	&	19.28	&	1.75	&	0.60	&	0.97	\\
Lm20810-4	&	-2	&	4	&	0.82	&	4.60	&	0.184	&	6.95	&	0.64	&	14.30	&	4.11	&	0.22	&	0.92	\\
Lm20810-5	&	-2	&	5	&	0.82	&	3.09	&	0.154	&	6.92	&	0.68	&	12.90	&	3.86	&	0.16	&	0.90	\\
Lm20810-6	&	-2	&	6	&	0.82	&	2.26	&	0.136	&	6.71	&	0.68	&	12.11	&	3.97	&	0.12	&	0.89	\\
Lm20810-7	&	-2	&	7	&	0.82	&	1.51	&	0.105	&	7.75	&	0.88	&	11.72	&	3.38	&	0.13	&	0.87	\\
Lm20810-8	&	-2	&	8	&	0.82	&	0.71	&	0.057	&	8.74	&	0.98	&	12.98	&	3.45	&	0.12	&	0.84	\\
Lm20810-9	&	-2	&	9	&	0.82	&	0.49	&	0.044	&	8.78	&	0.99	&	13.03	&	3.76	&	0.10	&	0.84	\\
Lm20810-10	&	-2	&	10	&	0.82	&	0.37	&	0.037	&	9.40	&	1.08	&	13.03	&	3.61	&	0.10	&	0.83	\\
\hline
Lm10810-1	&	-1	&	1	&	0.82	&	48.49	&	0.48	&	7.61	&	0.35	&	67.15	&	2.01	&	10.47	&	0.78	\\
Lm10810-2	&	-1	&	2	&	0.82	&	21.12	&	0.42	&	8.33	&	0.59	&	46.33	&	2.14	&	5.75	&	0.83	\\
Lm10810-3	&	-1	&	3	&	0.82	&	7.51	&	0.225	&	8.20	&	0.78	&	17.25	&	1.71	&	0.58	&	0.94	\\
Lm10810-4	&	-1	&	4	&	0.82	&	4.17	&	0.167	&	6.94	&	0.65	&	13.10	&	3.91	&	0.23	&	0.90	\\
Lm10810-5	&	-1	&	5	&	0.82	&	2.83	&	0.142	&	6.79	&	0.67	&	11.97	&	3.92	&	0.17	&	0.89	\\
Lm10810-6	&	-1	&	6	&	0.82	&	1.59	&	0.095	&	6.44	&	0.62	&	11.44	&	4.24	&	0.12	&	0.86	\\
Lm10810-7	&	-1	&	7	&	0.82	&	0.74	&	0.051	&	7.53	&	0.77	&	12.51	&	4.32	&	0.11	&	0.83	\\
Lm10810-8	&	-1	&	8	&	0.82	&	0.51	&	0.041	&	9.20	&	1.03	&	12.72	&	3.69	&	0.13	&	0.83	\\
Lm10810-9	&	-1	&	9	&	0.82	&	0.38	&	0.034	&	8.87	&	1.00	&	12.61	&	3.84	&	0.10	&	0.83	\\
Lm10810-10	&	-1	&	10	&	0.82	&	0.30	&	0.030	&	9.40	&	1.09	&	12.58	&	3.67	&	0.10	&	0.82	\\
\hline
L00810-1	&	0	&	1	&	0.82	&	48.57	&	0.49	&	8.43	&	0.53	&	67.09	&	2.01	&	10.61	&	0.78	\\
L00810-2	&	0	&	2	&	0.82	&	21.21	&	0.42	&	8.36	&	0.59	&	46.48	&	2.13	&	5.83	&	0.83	\\
L00810-3	&	0	&	3	&	0.82	&	7.21	&	0.216	&	8.15	&	0.78	&	16.52	&	1.70	&	0.57	&	0.93	\\
L00810-4	&	0	&	4	&	0.82	&	3.93	&	0.157	&	6.76	&	0.62	&	12.46	&	4.06	&	0.22	&	0.89	\\
L00810-5	&	0	&	5	&	0.82	&	0.94	&	0.047	&	7.46	&	0.70	&	12.27	&	4.23	&	0.15	&	0.82	\\
L00810-6	&	0	&	6	&	0.82	&	0.56	&	0.034	&	7.48	&	0.72	&	12.16	&	4.27	&	0.11	&	0.82	\\
L00810-7	&	0	&	7	&	0.82	&	0.34	&	0.024	&	8.79	&	0.97	&	12.17	&	3.77	&	0.10	&	0.82	\\
L00810-8	&	0	&	8	&	0.82	&	0.30	&	0.024	&	8.83	&	0.98	&	12.10	&	3.75	&	0.10	&	0.82	\\
L00810-9	&	0	&	9	&	0.82	&	0.24	&	0.022	&	9.38	&	1.08	&	12.06	&	3.59	&	0.10	&	0.82	\\
L00810-10	&	0	&	10	&	0.82	&	0.20	&	0.020	&	7.63	&	0.85	&	11.84	&	4.45	&	0.06	&	0.82	\\
\hline
\hline
Lm21010-4	&	-2	&	4	&	1.00	&	2.01	&	0.080	&	10.29	&	0.95	&	40.94	&	4.68	&	0.27	&	1.05	\\
Lm21010-5	&	-2	&	5	&	1.00	&	0.64	&	0.032	&	9.52	&	0.83	&	38.03	&	4.96	&	0.14	&	1.01	\\
Lm21010-6	&	-2	&	6	&	1.00	&	0.33	&	0.020	&	11.12	&	1.10	&	36.58	&	4.43	&	0.12	&	1.00	\\
Lm21010-7	&	-2	&	7	&	1.00	&	0.21	&	0.015	&	11.41	&	1.14	&	35.86	&	4.13	&	0.11	&	1.00	\\
Lm21010-8	&	-2	&	8	&	1.00	&	0.15	&	0.012	&	11.65	&	1.17	&	35.50	&	3.89	&	0.11	&	1.00	\\
Lm21010-9	&	-2	&	9	&	1.00	&	0.12	&	0.010	&	10.92	&	1.10	&	35.14	&	4.48	&	0.08	&	1.00	\\
Lm21010-10	&	-2	&	10	&	1.00	&	0.09	&	0.009	&	11.13	&	1.14	&	34.97	&	4.46	&	0.07	&	1.00	\\
\hline
Lm11010-4	&	-1	&	4	&	1.00	&	0.89	&	0.035	&	9.42	&	0.80	&	37.40	&	5.17	&	0.21	&	1.01	\\
Lm11010-5	&	-1	&	5	&	1.00	&	0.41	&	0.020	&	10.91	&	1.07	&	35.68	&	4.74	&	0.14	&	1.00	\\
Lm11010-6	&	-1	&	6	&	1.00	&	0.24	&	0.015	&	10.48	&	1.00	&	34.93	&	4.41	&	0.12	&	1.00	\\
Lm11010-7	&	-1	&	7	&	1.00	&	0.17	&	0.012	&	11.15	&	1.10	&	34.58	&	4.00	&	0.11	&	1.00	\\
Lm11010-8	&	-1	&	8	&	1.00	&	0.12	&	0.010	&	10.60	&	1.06	&	34.24	&	4.49	&	0.08	&	1.00	\\
Lm11010-9	&	-1	&	9	&	1.00	&	0.10	&	0.009	&	10.76	&	1.09	&	34.06	&	4.51	&	0.07	&	1.00	\\
Lm11010-10	&	-1	&	10	&	1.00	&	0.08	&	0.008	&	11.27	&	1.18	&	33.94	&	4.50	&	0.06	&	1.00	\\
\hline
L01010-4	&	0	&	4	&	1.00	&	0.35	&	0.014	&	9.56	&	0.86	&	33.79	&	4.53	&	0.12	&	1.00	\\
L01010-5	&	0	&	5	&	1.00	&	0.21	&	0.010	&	10.31	&	0.97	&	33.32	&	3.91	&	0.11	&	1.00	\\
L01010-6	&	0	&	6	&	1.00	&	0.15	&	0.009	&	9.85	&	0.95	&	32.95	&	4.47	&	0.07	&	1.00	\\
L01010-7	&	0	&	7	&	1.00	&	0.11	&	0.008	&	10.85	&	1.13	&	32.77	&	4.48	&	0.06	&	1.00	\\
L01010-8	&	0	&	8	&	1.00	&	0.09	&	0.007	&	11.94	&	1.31	&	32.63	&	4.52	&	0.05	&	1.00	\\
L01010-9	&	0	&	9	&	1.00	&	0.07	&	0.006	&	12.84	&	1.44	&	32.54	&	4.50	&	0.05	&	1.00	\\
L01010-10	&	0	&	10	&	1.00	&	0.06	&	0.006	&	13.43	&	1.52	&	32.51	&	4.47	&	0.05	&	1.00	\\
\hline
\hline
Lm11110-4	&	-1	&	4	&	1.10	&	0.35	&	0.014	&	11.20	&	0.95	&	73.91	&	5.45	&	0.15	&	1.11	\\
Lm11110-5	&	-1	&	5	&	1.10	&	0.20	&	0.010	&	11.72	&	1.01	&	72.65	&	4.60	&	0.14	&	1.10	\\
Lm11110-6	&	-1	&	6	&	1.10	&	0.13	&	0.008	&	16.78	&	1.76	&	71.27	&	5.99	&	0.07	&	1.10	\\
Lm11110-7	&	-1	&	7	&	1.10	&	0.10	&	0.007	&	19.61	&	2.05	&	70.91	&	6.00	&	0.06	&	1.10	\\
Lm11110-8	&	-1	&	8	&	1.10	&	0.07	&	0.006	&	24.90	&	2.51	&	70.41	&	6.00	&	0.05	&	1.10	\\
Lm11110-9	&	-1	&	9	&	1.10	&	0.06	&	0.005	&	28.76	&	2.79	&	70.18	&	6.00	&	0.04	&	1.10	\\
Lm11110-10	&	-1	&	10	&	1.10	&	0.05	&	0.005	&	32.16	&	3.01	&	70.00	&	6.00	&	0.04	&	1.10	\\
\hline
L01110-4	&	0	&	4	&	1.10	&	0.20	&	0.008	&	10.67	&	0.87	&	68.70	&	4.64	&	0.12	&	1.10	\\
L01110-5	&	0	&	5	&	1.10	&	0.13	&	0.006	&	19.38	&	2.05	&	67.72	&	6.00	&	0.06	&	1.10	\\
L01110-6	&	0	&	6	&	1.10	&	0.09	&	0.006	&	23.09	&	2.38	&	67.38	&	6.00	&	0.05	&	1.10	\\
L01110-7	&	0	&	7	&	1.10	&	0.07	&	0.005	&	27.35	&	2.71	&	67.12	&	6.00	&	0.04	&	1.10	\\
L01110-8	&	0	&	8	&	1.10	&	0.06	&	0.004	&	32.29	&	3.04	&	66.90	&	6.00	&	0.03	&	1.10	\\
L01110-9	&	0	&	9	&	1.10	&	0.05	&	0.004	&	36.19	&	3.28	&	66.73	&	6.00	&	0.03	&	1.10	\\
L01110-10	&	0	&	10	&	1.10	&	0.04	&	0.004	&	39.39	&	3.47	&	66.59	&	6.00	&	0.02	&	1.10	\\
\hline
\end{longtable}
\end{longtab}

\subsection{Rotational profile} \label{sec:rotational-profile}
\begin{figure}
   \centering
   \begin{subfigure}[b]{1.0\textwidth}
   \input{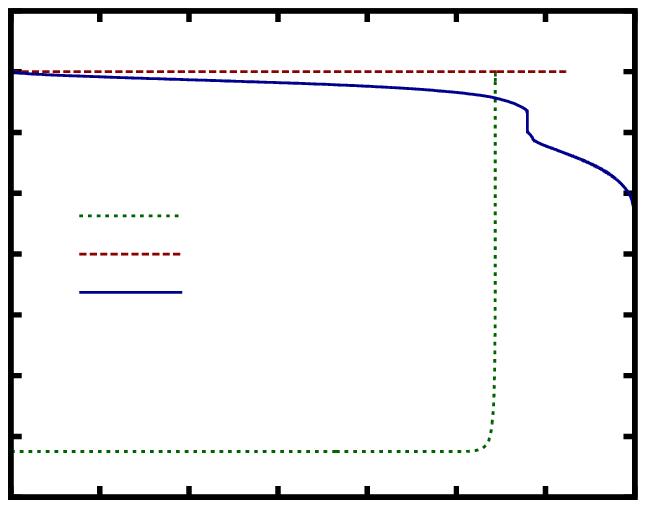}
   \end{subfigure} 
   \begin{subfigure}[t]{1.0\textwidth}
   \input{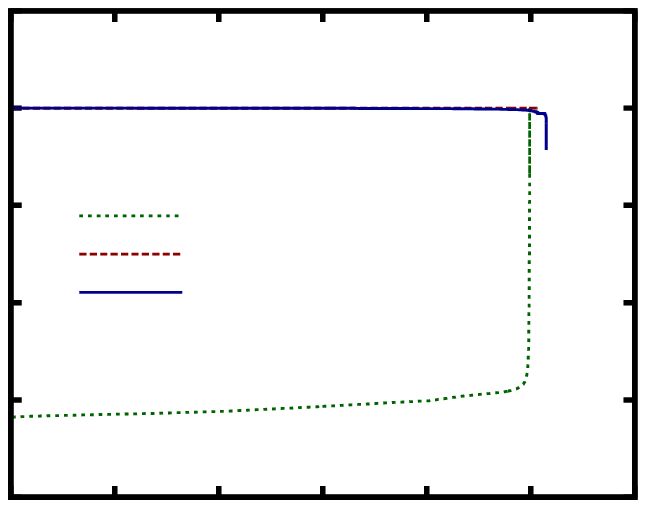}
   \end{subfigure}
   \normalsize
      \caption{Rotational velocity profile of model sequence \mbox{Lm20510-5} ($M_\mathrm{init}=0.54~\mathrm{M}_\odot$) (A) and \mbox{Lm21010-5} ($M_\mathrm{init}=1.0~\mathrm{M}_\odot$) (B) at three different times during its evolution ($f_\mathrm{acc}=1.0$, $\dot{M}=5\cdot10^{-8}~\mathrm{M}_\odot/\mathrm{yr}$ and $\log{(L_\mathrm{init}/\mathrm{L}_\odot)}=-2$ in both cases). Line i corresponds to $t=0~\mathrm{yr}$, the start of mass accretion, m to the midpoint of the accretion phase and f to just after the point of He-ignition. In subfigure (A) this corresponds to m at $t\sim1.55~\mathrm{Myr}$ and f at $t=3.11~\mathrm{Myr}$. In subfigure (B) the corresponding times are m at $t\sim0.3~\mathrm{Myr}$ and f at $t=0.64~\mathrm{Myr}$. The y-axis is normalized to unity with the maximum value of each graph indicated in $\omega_\mathrm{max}$. }
         \label{fig:radv}
\end{figure}
WDs tend to be convectively stable throughout their interior. Accordingly, the Tayler-Spruit mechanism will work on the entirety of their structure. When compared to dynamical shears, the strength of their associated shear viscosities is of the same magnitude for a given condition, but the critical rotational velocity gradient required to initiate the Tayler-Spruit mechanism (Eq.~\ref{eq:minq}) is usually much smaller. Without angular momentum accretion, the dynamo will remain active until the velocity gradient is decreased to that of the threshold gradient.

Fig.~\ref{fig:radv}A shows the radial velocity profile of model sequence \mbox{Lm20510-5}\footnote{The model number is composed as follows: The first digit indicates the value of the initial luminosity, if preceded by the letter "m", that value should be understood to be negative. The second and third digit is equal to ten times the value of the the initial mass rounded to the next $0.1~\mathrm{M}_\odot$. The fourth and fifth digit are equal to the value of ten times the chosen value for $f$. The hyphenated number is equal to the value of the chosen mass accretion rate divided by $10^{-8}$.}. As can be seen, any residual differential rotation present at the beginning of mass accretion (line i in Fig.~\ref{fig:radv}A) is damped by the Tayler-Spruit machanism until a situation close to solid body rotation is attained (line m). A very small radial velocity gradient remains, but is barely noticeable in the plot. Line f in Fig.~\ref{fig:radv}A shows the profile just after ($\ll 1~\mathrm{s}$) He-ignition. With the onset of rapid expansion, conservation of angular momentum in the expanding layers leads to the reemergence of an angular velocity gradient. The situation is similar for more massive WDs, as shown in Fig.~\ref{fig:radv}B.

Note that there is no significant change of the rotational profile even as it passes areas with significant gradient in the chemical composition. Such a change might be expected due to the dependence of the effective viscosity on the chemical stratification (see Eq.~\ref{eq:compn}).

\begin{figure}
   \centering
   \begin{subfigure}[b]{1.0\textwidth}
   \input{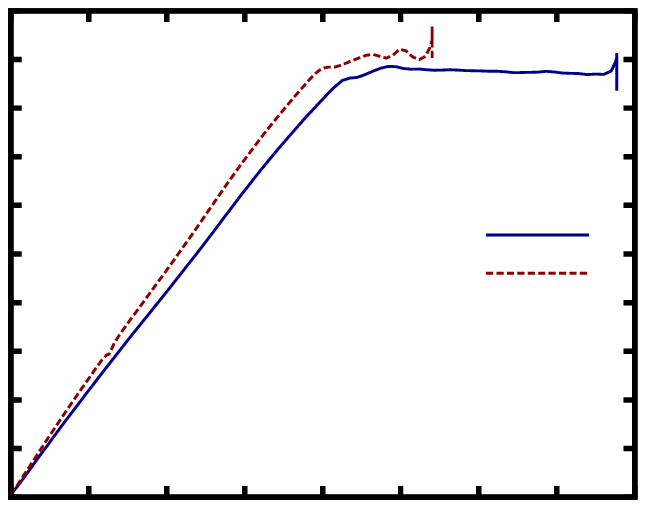}
   \end{subfigure} 
   \begin{subfigure}[t]{1.0\textwidth}
   \input{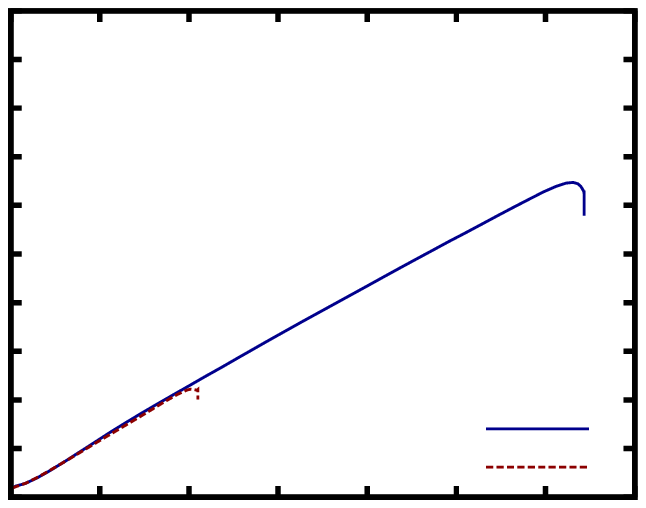}
   \end{subfigure}
   \normalsize
      \caption{Evolution of the surface velocities of model sequences L00510-5, Lm20510-5 (A) and L01010-5, Lm21010-5 (B). All model sequences are computed up to He-ignition. The y-axis is normalized to the critical rotational velocity. $M_\mathrm{env}$ is the mass of the helium envelope and corresponds to the amount of helium accreted. All models were computed with $\dot{M}=5\cdot10^{-8}~\mathrm{M}_\odot/\mathrm{yr}$ and $f_\mathrm{acc}=1.0$.}
         \label{fig:wcritm}
\end{figure}

Fig.~\ref{fig:wcritm} shows the surface velocity of four representative model sequences at two different initial luminosities of $\log{L_\mathrm{init}/\mathrm{L}_\odot}=0$ and $\log{L_\mathrm{init}/\mathrm{L}_\odot}=-2$ (\mbox{L00510-5}, \mbox{Lm20510-5}A and \mbox{L01010-5}, \mbox{Lm21010-5}B) in units of $\omega_\mathrm{surf}/\omega_\mathrm{crit}$ ($\omega_\mathrm{crit}$ being the critical surface velocity) with respect to time. Less massive WDs spin up faster than more massive ones (See Fig.~\ref{fig:wcritm}) due to their relatively smaller rotational inertia. The fact that the models with $M_\mathrm{init}=0.54~\mathrm{M}_\odot$ never quite reach $\omega_\mathrm{crit}$ is related to the treatment of super-critical accretion as explained in Sec.~\ref{sec:methods}. The prescription calls for $f_\mathrm{acc}=0$ if further accretion would lead to super-critical rotation, which means that critical rotation itself can never actually be reached.  

Fig.~\ref{fig:wcritm} further shows that the fraction of the critical rotational velocity achievable by a WD strongly depends on the initial mass of each model. The reasons for this, while somewhat obvious, as well as the implications of this behavior will be discussed in greater detail in Sec.~\ref{sec:ignitions}. It should be noted, though, that none of our model sequences of $M_\mathrm{init}>0.82~\mathrm{M}_\odot$ reach critical rotation before He-ignition, which impacts the mass-dependency of final He-shell masses, as will be discussed in Sec.~\ref{sec:ignitions}.

\subsection{Chemical profile} \label{sec:chemical-profile}

\begin{figure}
   \centering
   \small
      \begin{subfigure}[b]{1.0\textwidth}
   \input{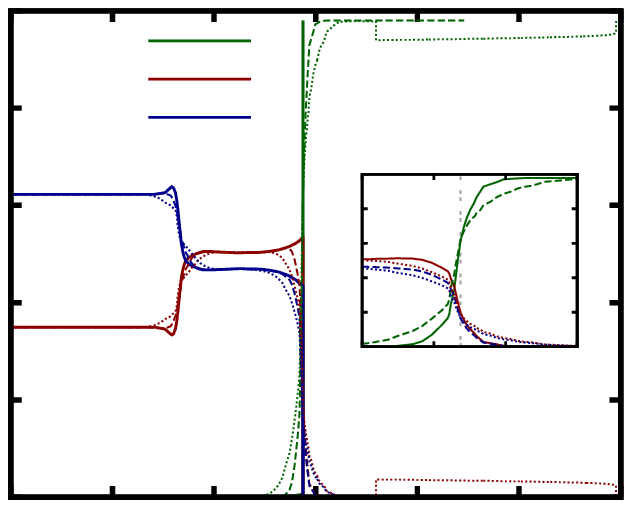}
   \end{subfigure} 
   \begin{subfigure}[t]{1.0\textwidth}
   \input{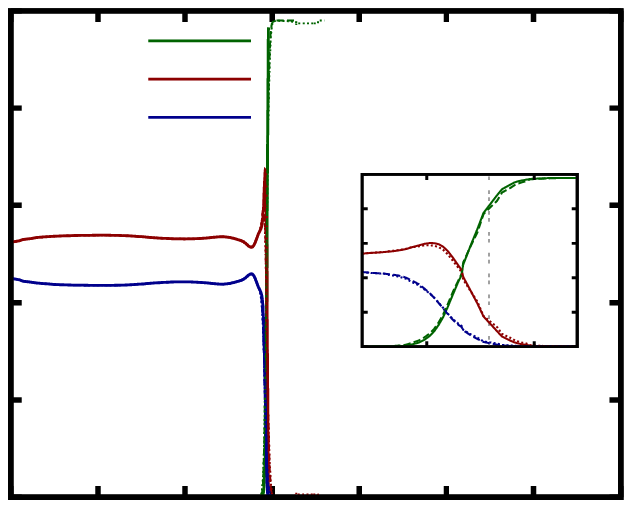}
   \end{subfigure}
   \normalsize
      \caption{Chemical profile of model sequence \mbox{Lm20510-5} ($M_\mathrm{init}=0.54~\mathrm{M}_\odot$) (A) and \mbox{Lm21010-5} ($M_\mathrm{init}=1.0~\mathrm{M}_\odot$) (B). The remaining initial parameters are the same for both model sequences: $f_\mathrm{acc}=1.0$, $\dot{M}=5\cdot10^{-8}~\mathrm{M}_\odot/\mathrm{yr}$ and $\log{(L_\mathrm{init}/\mathrm{L}_\odot)}=-2$. Included are the three most abundant isotopes $\mathrm{He}^4$, $\mathrm{C}^{12}$ and $\mathrm{O}^{16}$ in units of relative mass per shell over the shell's mass coordinate. Solid lines represent the profile at the time of the start of mass accretion at $t=0~\mathrm{yr}$, dashed lines at the midpoint of the accretion phase at $t\sim1.55~\mathrm{Myr}$ (A) and $t\sim0.3~\mathrm{Myr}$ (B) dotted lines just after the point of He-ignition at $t=3.11~\mathrm{Myr}$ (A) and $t=0.64~\mathrm{Myr}$ (B). The inset is a zoom of the area of the initial envelope-core interface (the interface is shown as a dashed, grey line), where, for the sake of clarity, the initial profile has been omitted.}
         \label{fig:comp}
\end{figure}

Fig.~\ref{fig:comp} shows the chemical development of two representative model sequences, \mbox{Lm20510-5} and \mbox{Lm21010-5}. As can be seen, there is noticeable chemical mixing at the boundary layer of the less massive system (Fig.~\ref{fig:comp} A), while the more massive system (Fig.~\ref{fig:comp} B) shows barely any mixing at the boundary layer. Rotationally induced mixing has been found to be much more pronounced in non-magnetic rotating models \citep{YL2004b}. With the Tayler-Spruit mechanism suppressing the overwhelming majority of any initially present angular velocity gradient, the only region where appreciable chemical mixing can plausibly take place is the outermost layer of the envelope where an angular velocity gradient is maintained by the continuous accumulation of rotating matter. With growing envelope mass, the area exhibiting the largest angular velocity gradient moves away from the initial surface mass coordinate. This means that with growing envelope mass, chemical mixing at the core-envelope interface gradually diminishes. The differences in the mixing behavior between the two presented model sequences are due to a difference in initial chemical profiles.

We therefore conclude that chemical mixing in these systems is fundamentally weak, but highly dependent on the initial chemical profile. The presence of a preexisting He-envelope considerably weakens the magnitude of expected chemical diffusion.

The expectation of an adequately well-defined interface between the original core and the accumulated material throughout the evolution of our WD models leads us to refer to this interface as the "core-envelope interface" from this point on.

\subsection{Accretion behaviour and ignition} \label{sec:ignitions}

\begin{figure}
   \centering
   \small
   \input{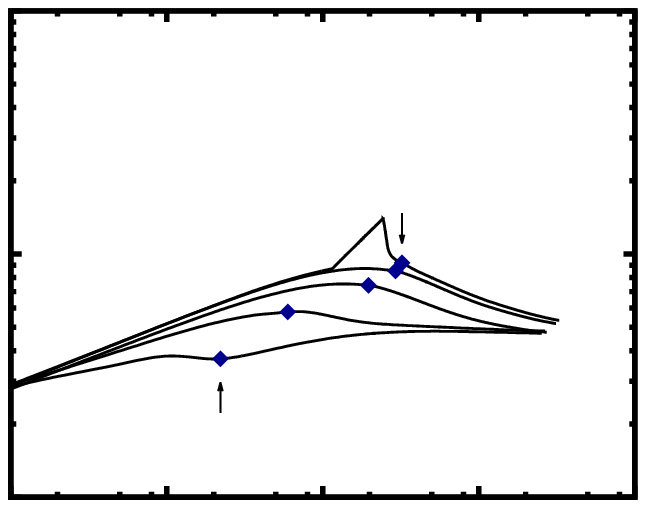}
   \input{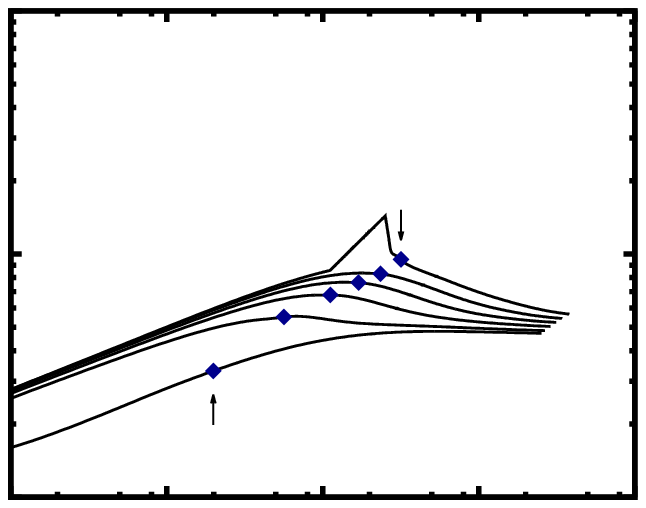}
   \normalsize
      \caption{Evolution of the temperature-density profile of a rotating (A) and non-rotating (B) WD with $0.54~\mathrm{M}_\odot$ accretion rate is $5 \cdot 10^{-8}~\mathrm{M}_\odot/\mathrm{yr}$, $f_\mathrm{acc}=1.0$ and $\log{(L_\mathrm{init}/\mathrm{L}_\odot)}=-1$. Blue dots indicate the core-envelope interface. Initial (at the beginning of mass accretion) and final (at the point of helium ignition) profiles are indicated. The given masses (left to right) correspond to the presented profiles (initial to final). Note that the profiles marked as "initial" are not the first models in their evolutionary sequence, having been evolved for about one kyr each.}
         \label{fig:ign06}
\end{figure}

\begin{figure}
   \centering
   \small
   \input{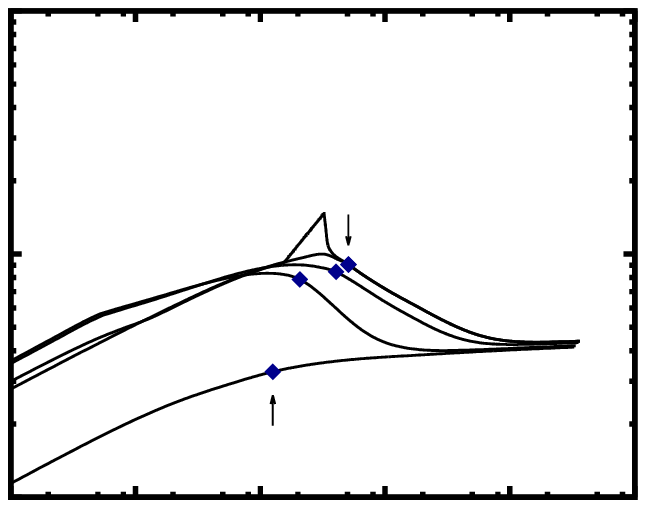}
   \input{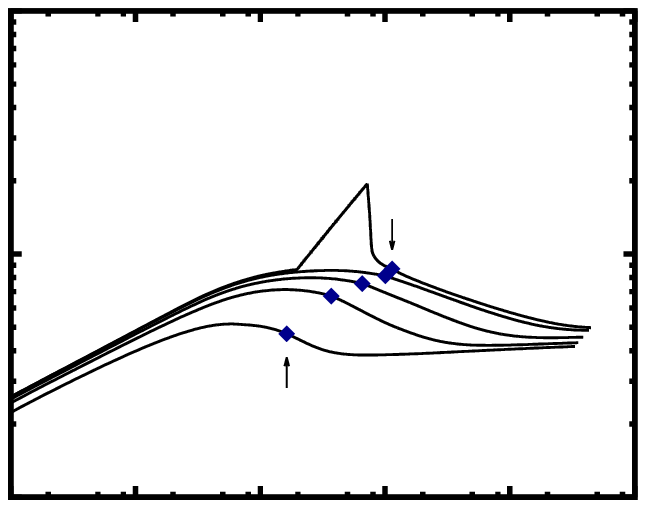}
   \normalsize
      \caption{Evolution of the temperature-density profile of a rotating (A) and non-rotating (B) WD with $1.0~\mathrm{M}_\odot$. All other parameters and labels as in Fig.~\ref{fig:ign06}.}
         \label{fig:ign10}
\end{figure}

\begin{figure*}[]
   \centering
   \small
   \input{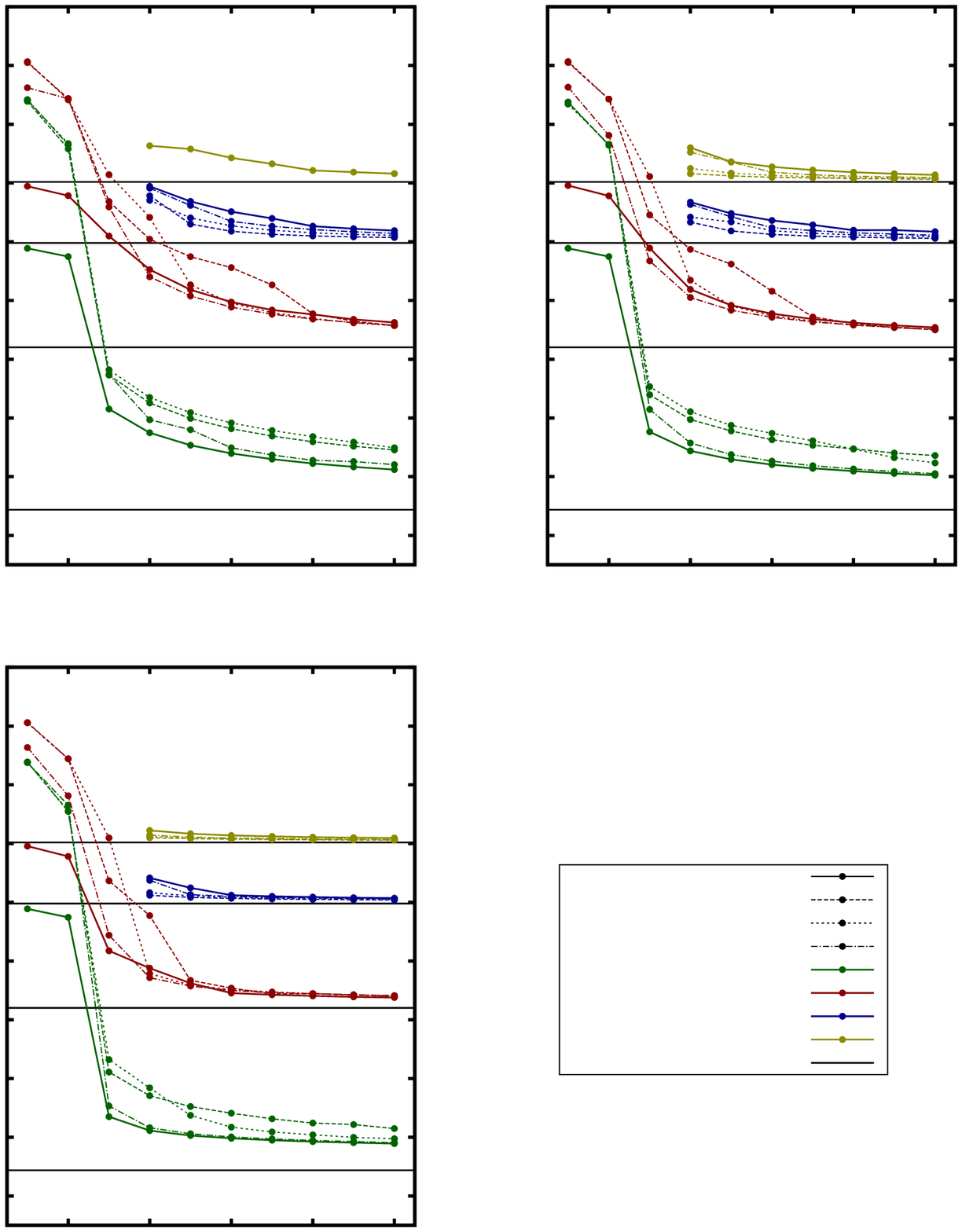}
   \normalsize
      \caption{Total masses ($M$) of all computed model sequences of this study with respect to constant mass accretion rate ($\dot{M}$). The presented graphs differ in the chosen initial luminosity of the white dwarf model with $\log{(L_\mathrm{init}/\mathrm{L}_\odot)}=-2$ in graph (A), $\log{(L_\mathrm{init}/\mathrm{L}_\odot)}=-1$ in (B) and $\log{(L_\mathrm{init}/\mathrm{L}_\odot)}=0$ in (C). Black solid lines correspond to the initial mass of each model sequence with the initial mass being indicated. Colored lines indicate the final mass of each model sequence with green corresponding to $M_\mathrm{init}=0.54~\mathrm{M}_\odot$, red to $M_\mathrm{init}=0.82~\mathrm{M}_\odot$, blue to $M_\mathrm{init}=1.00~\mathrm{M}_\odot$ and yellow to $M_\mathrm{init}=1.10~\mathrm{M}_\odot$. Line types indicate chosen rotational parameters. Solid colored lines for non-rotating models, dashed lines for $f_\mathrm{acc}=1.0$, dashed lines for $f_\mathrm{acc}=0.6$ and dash-dotted lines for $f_\mathrm{acc}=0.3$. Dots indicate the masses of individual models.} 
         \label{fig:fmasses}
\end{figure*}

Figs.~\ref{fig:ign06} and \ref{fig:ign10} show the evolution of representative model sequences in the $\rho$-$T$-plane. The model sequence in Fig.~\ref{fig:ign06}A depicts model sequence \mbox{Lm10510-5}, which means that $M_\mathrm{init}=0.54~\mathrm{M}_\odot$, $\log{L_\mathrm{init}/\mathrm{L}_\odot}=-1$ and $f_\mathrm{acc}=1.0$. Fig.~\ref{fig:ign06}B is the same but non-rotating. The initial profile is as at the beginning of the mass accretion phase and the final one is just before ($\leq 1~\mathrm{s}$) He-ignition as defined in Sec.~\ref{sec:methods}. It should be noted, however, that the rotating model accumulated a significantly larger amount ($0.05~\mathrm{M}_\odot$) of helium at the point of detonation. This is a consequence of solid body rotation. Helium ignition requires both sufficiently high temperatures and sufficiently high densities, the threshold temperature being a function of local density. In the rotating, but non-magnetic case, the WD would, at the point of He-ignition, show significant differential rotation, introducing a much larger amount of energy dissipation (Eq. \ref{eq:ediss}) into the uppermost layers of the He-envelope, thereby providing the necessary temperature to facilitate He-ignition after a much smaller amount of helium has been accumulated \citep[see][]{YL2004a}. The Tayler-Spruit mechanism is active at much smaller rotational velocity gradients than, e.g. the dynamical shear instability, under the conditions encountered in the present model sequences. This means that, as described in Sec.~\ref{sec:rotational-profile}, a small rotational velocity gradient is induced which extends throughout the stellar envelope (i.e. a small, non-localized gradient). The energy dissipation caused by the Tayler-Spruit mechanism is correspondingly small (locally, practically $\epsilon_\mathrm{diss} \approx 0$). This means that a temperature increase resulting from energy dissipation due to the Tayler-Spruit mechanism will be smaller but affect a wider area when compared to that expected from, e.g. the dynamical shear instability. Therefore, in order to increase the temperature of the stellar material at any given point sufficiently to induce He-ignition, a larger amount of angular momentum, and thus helium, has to be accreted than in the rotating non-magnetic case. Further, since the WD is rotating critically (see Fig.~\ref{fig:wcritm} A, this model sequence falls between the two depicted ones), the centrifugal force serves to decrease the local density of the stellar matter at any given coordinate compared to the non-rotating case, in turn increasing the temperature required for He-ignition. We therefore expect a significant increase in the required amount of material to be accreted before ignition. It should be noted at this point that the evolution of the rotational profile is the dominant factor in this prediction. This becomes evident when comparing the depicted model sequences with models computed with a variation of values of the $f_\mathrm{acc}$-parameter. One finds that, $M_\mathrm{init}$ held constant at $0.54~\mathrm{M}_\odot$, at high accretion rates, the final accumulated amount of helium decreases with increasing $f_\mathrm{acc}$ (see Fig.~\ref{fig:fmasses}). Exceptions to this exist at low accretion rates. These exceptions are discussed in detail in Sec.~\ref{sec:f-dependence}.

Fig.~\ref{fig:ign10} is the same as Fig.~\ref{fig:ign06} but the model sequence's initial mass is $M_\mathrm{init}=1.00~\mathrm{M}_\odot$. Unlike in the less massive example, the effects of rotation lead to a decrease of the required mass to be accreted in order to initiate He-ignition. Partially this is a consequence of the physical differences in WDs of different mass at the onset of mass accretion (see Tab.~\ref{tab:inmodel}). Smaller size at a higher mass makes newly deposited material move at much higher angular velocity for a given value of $f_\mathrm{acc}$. This leads to larger amounts of dissipational energy being introduced in the system. This combined with the compressional heating incurred by the accumulating material falling onto the WD with a significantly higher surface gravity than in less massive cases, leads to a much faster increase in temperature in the outer layers of the WD (compare Figs.~\ref{fig:ign06}A and \ref{fig:ign10}A). 

Rotational inertia of white dwarfs increases with increasing mass \citep[see][for example]{PSIA2003}. The higher the rotational inertia of a solidly or quasi-solidly rotating object, the smaller the increase of rotational velocity incurred through accretion of a given amount of rotational momentum. This leads to He ignition occurring before the WD has a chance to achieve critical or near-critical rotation (see Fig.~\ref{fig:wcritm}), in turn leading to much diminished centrifugal effects compared to the case of lower masses.

In both of the cases presented above, it is useful to comment on the NCO reaction. The effects of the NCO reaction on accretion scenarios have been investigated before \citep[e.g.][]{WW1994, HNAK1986}. The consensus is that the reaction is largely, but not completely \citep{HNAK1986}, independent of temperature, the dominant factors in a CO WD being a sufficiently high local abundance of nitrogen and a local density of more than $10^6~\mathrm{g/cm}^3$. In the present scenario, in the absence of strong chemical mixing (Sec.~\ref{sec:chemical-profile}), the effect of increased energy generation due to the NCO reaction would be most important at the core-envelope interface. In both the rotating and non-rotating model sequences with $M_\mathrm{init}=0.54~\mathrm{M}_\odot$, the interface never reaches densities large enough to facilitate the NCO reaction. In the heavier model sequences, only the non-rotating WD exhibits high enough densities at the interface, leading to an acceleration of the temperature increase due to the NCO reaction in the overlying layers and therefore earlier He-ignition than if the NCO reaction were neglected. In our model sequences the NCO-reaction plays a limited role insofar as it provides some additional heat in models with low initial luminosities. It only leads to a slightly reduced final helium envelope mass and to the igniting layer being located somewhat lower in the star. The majority of the heat is still introduced through the accretion process itself. In models with higher initial luminosities, the role of the NCO-reaction is eclipsed by the residual heat of the WD.

The answer to the question whether rotating magnetic WDs accumulate more massive He-envelope at He-ignition than non-rotating ones therefore depends strongly on whether the WD can achieve near-critical rotation before He-ignition and therefore on the initial mass as well as the efficiency of angular momentum accumulation.

\begin{figure}
   \centering
   \small
   \input{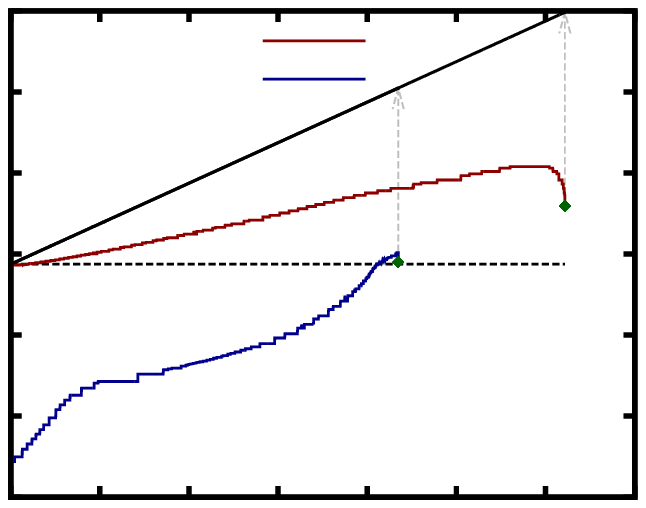}
   \input{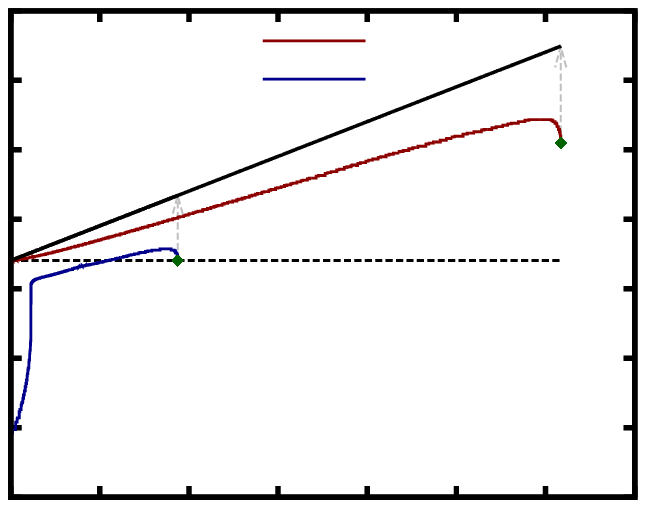}
   \input{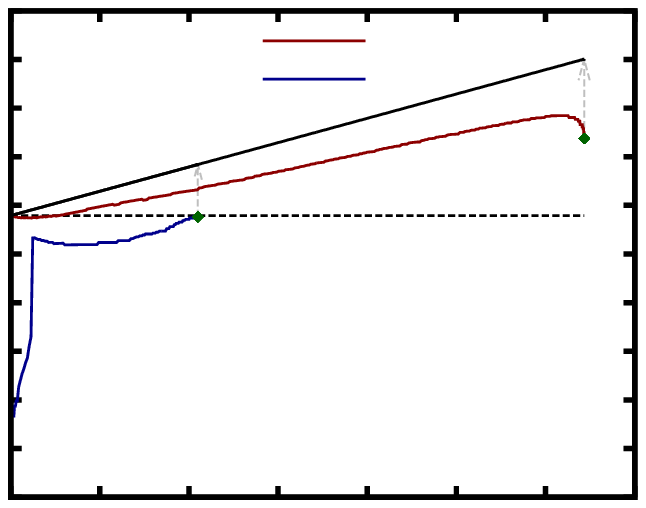}
   \normalsize
      \caption{Mass coordinates of the point of the temperature maximum ($T_\mathrm{max}$) over time (solid colored lines) of several representative systems with initial masses of $0.54~\mathrm{M}_\odot$ (A), $0.82~\mathrm{M}_\odot$ (B) and $1.0~\mathrm{M}_\odot$ (C). The solid black line in each plot indicates the corresponding total mass of the WD ($M_\mathrm{WD}$), red lines correspond to an initial luminosity of $\log(L_\mathrm{init}/\mathrm{L}_\odot)=0$ and blue lines to $\log(L_\mathrm{init}/\mathrm{L}_\odot)=-2$. The dashed black line indicates the initial mass of the WD. The green dots indicate the He-ignition points. All depicted sequences were calculated with $f_\mathrm{acc}=1.0$ and $\dot{M}=5 \cdot 10^{-8} \mathrm{M}_\odot/\mathrm{yr}$. The grey arrows indicate the point of ignition in relation to $M_\mathrm{WD}$. $M_\mathrm{r,ign}$ is the mass coordinate of the igniting shell.}
         \label{fig:maxtla}
\end{figure} 
Fig.~\ref{fig:maxtla} shows the time-evolution of the location of the maximum temperature in the star ($T_\mathrm{max}$) of several representative model sequences at different initial masses ($0.54~\mathrm{M}_\odot$ A, $0.82~\mathrm{M}_\odot$ B and $1.0~\mathrm{M}_\odot$ C) with two initial luminosities of $\log(L_\mathrm{init}/\mathrm{L}_\odot)=0$ and $\log(L_\mathrm{init}/\mathrm{L}_\odot)=-2$, calculated with $f_\mathrm{acc}=1.0$. 

Since the initial temperature of WDs of the same mass is, across the entire interior, significantly higher at higher initial luminosities, the tendency of $M(T_\mathrm{max})$ to stay at lower values in models with higher initial luminosities is expected. If the core temperature is higher, a larger amount of thermal energy has to be introduced into the star in order to move $M(T_\mathrm{max})$ towards the surface, albeit aided by extant heat in the core.

It is easily seen that the mass coordinate of He-ignition strongly depends on the initial luminosity of the WD. Higher initial luminosities cause the point of ignition to lie closer to the core-envelope interface (the presented model sequences at $\log(L_\mathrm{init}/\mathrm{L}_\odot)=0$ feature ignition coordinates not significantly above the core-envelope interface). As additional heat is introduced close to the WD's surface, $M(T_\mathrm{max})$ moves towards the surface, but higher initial luminosities imply a higher core temperature, causing the $M(T_\mathrm{max})$ to remain closer to the center of the WD. The visible downturn of $M(T_\mathrm{max})$ in each of the depicted model sequences is because of the burning point moving towards areas of higher density after ignition.

\cite{WK2011} investigated the two possible ways to induce SNe in sub-Chandrasekhar WDs via what has become known as the double-detonation scenario in closer detail: In the first scenario, ignition of the He-envelope at the core-envelope interface induces a supersonic shock front which compresses the CO-core, inducing a secondary detonation at the center and resulting in the destruction of the WD and a SN. In the second scenario, an ignition at some distance the core-envelope interface induces an inward and outward shock. The inward propagating shock then proceeds from the Helium envelope and into the CO core and induces a SN ("edge-lit scenario"). In the light of this study, assuming that an ignition is able to induce a detonation, the former case would be more likely in an initially bright WD and the latter more likely in an initially faint one.

\subsection{Impact of parameter variations} \label{sec:variations}

In this section we describe the impact of variations of the most relevant input parameters on the entire set of simulated systems.

Fig.~\ref{fig:fmasses} shows the final masses of all computed model sequences with respect to the mass accretion rate. We emphasize that the coordinate in the presented graphs does not correspond to the mass of the accumulated envelope but to the total mass of the star. The mass of the envelope can be obtained by subtracting the mass of the corresponding initial model (black lines) from the mass of the model sequence at He-ignition.
We identify two distinct cases for accretion rates with $\dot{M}\geq 3 \cdot 10^{-8}~\mathrm{M}_\odot/\mathrm{yr}$ (effects at accretion rates lower than this, applicable to initial masses $M\geq 0.82~\mathrm{M}_\odot$ will be discussed separately), differentiated by their ignition mass when compared to the non-rotating case: The \textit{high initial mass case} in which the WD ignites its helium envelope after accumulation of a smaller amount of helium than the non-rotating case and the \textit{low initial mass case} in which the WD ignites its envelope with a larger amount of helium than in the non-rotating case. There is also the \textit{intermediate} case where the behavior of the WD is principally governed by the accretion rate. Systems with accretion rates lower than $\dot{M}<3\cdot10^{-8}~\mathrm{M}_\odot/\mathrm{yr}$ form a separate case, exhibiting vastly increased He-shell masses at detonation when compared to the non-rotating case as well as higher accretion rates. These systems will be collectively referred to as the \textit{low accretion rate case}.
\subsubsection{Low initial mass case} \label{sec:low-case}
The behavior of the low initial mass case is illustrated by the model sequences with $M_\mathrm{init}=0.54~\mathrm{M}_\odot$.
The effects of close-to-critical rotation, discussed in Sec.~\ref{sec:rotational-profile} and Sec.~\ref{sec:ignitions}, driving up ignition masses become obvious here. At all initial luminosities the model sequences with $M_\mathrm{init}=0.54~\mathrm{M}_\odot$ exhibit significantly higher envelope masses at He-ignition than in the non-rotating case (by up to $\sim0.14~\mathrm{M}_\odot$). The observation that this behavior is related to the ability of the accumulated angular momentum to spin up the WD is strongly supported by the evident correlation between the increase of the required accretion and $f_\mathrm{acc}$, since a variation in $f_\mathrm{acc}$ immediately and artificially impacts the amount of angular momentum imparted on the star for a given time period and therefore directly controls its spin-up timescale.
The switch (with $f_\mathrm{acc}=0.6$ showing higher ignition masses than $f_\mathrm{acc}=1.0$) at $\dot{M}\simeq4-5\cdot10^{-8}~\mathrm{M}_\odot/\mathrm{yr}$ in this behavior seen in Fig.~\ref{fig:fmasses} should be noted for further reference (see explanation in Sec.~\ref{sec:f-dependence}).

Systems with $f_\mathrm{acc}=0.3$ generally do not spin up quickly enough for the WD to reach surface velocities close to $v_\mathrm{crit}$, usually undergoing He-ignition at $v_\mathrm{surf} \sim 0.25~v_\mathrm{crit}$. These systems therefore, conforming to expectation, most closely resemble non-rotating ones, with exceptions at the lowest considered mass accretion rates if $\log{(L_\mathrm{init}/\mathrm{L}_\odot)}\leq-1$.

At $\log{(L_\mathrm{init}/\mathrm{L}_\odot)}=-2$, the initial temperatures are low enough, ensuring that the amount of required material is high enough to allow for a significant increase in rotational velocity, to ensure that the process described above is active over the entire range of considered mass accretion rates.

We conclude that the rotating magnetic case leads to higher envelope masses in relatively low mass CO WDs of about $0.54~\mathrm{M}_\odot$ compared to the non-rotating case.
\subsubsection{High initial mass case} \label{sec:high-case}
The behavior of the high initial mass case is illustrated by the model sequences with initial masses of $M_\mathrm{init}=1.00~\mathrm{M}_\odot$ and $M_\mathrm{init}=1.10~\mathrm{M}_\odot$. Unlike the low initial mass case, WDs of this kind undergo He-ignition after a smaller amount of material has been accumulated than in the non-rotating case. As described in Sec.~\ref{sec:rotational-profile} and Sec.~\ref{sec:ignitions}, the reason for this lies in the expectation, due to their higher rotational inertia, that higher mass WDs are spun up more slowly than lower mass objects. This means that higher mass WDs will not be spun up appreciably before He-ignition. The difference in ignition masses with respect to the value of $f_\mathrm{acc}$ in this case stems from the more efficient heating caused by rotational energy dissipation because of the larger rotational velocity gradient induced through accretion at larger values of $f_\mathrm{acc}$. This is reflected in the ordering of the ignition masses with the divergence from the non-rotating case being smallest for $f_\mathrm{acc}=0.3$ and largest for $f_\mathrm{acc}=1.0$. It should be noted that the decrease of ignition masses in the high initial mass case in the rotating magnetic case mirrors the behavior of rotating non-magnetic ones \cite[compare][]{YL2004b}. The non-rotating case and the rotating non-magnetic case can be thought of as upper and lower limits for the He-envelope mass at ignition respectively for the high initial mass case ($M\geq1.0~\mathrm{M}_\odot$).
\subsubsection{Intermediate case} \label{sec:intermediate-case}
The intermediate case is represented in this study by model sequences with $M_\mathrm{init}=0.82~\mathrm{M}_\odot$. These systems are characterized by exhibiting the behavior of high initial mass case systems at high mass transfer rates and low values of the $f_\mathrm{acc}$-parameter and the behavior of low initial mass case systems at low mass transfer rates and high values of the $f_\mathrm{acc}$-parameter. Lower initial luminosity increases the propensity of the system to show an increased He-envelope mass at ignition when compared to the non-rotating case. The reason for this behavior again lies in the interplay between the spin-up of the WD and the heating effects of accretion: Higher mass accretion rates decrease the amount of helium needed to result in an ignition, allowing less time to spin up the WD. The intermediate case can therefore be viewed as a transitory case between the high and low initial mass cases.
We again emphasize the expectation of the largest helium envelope masses to be reached to pass from model sequences with $f_\mathrm{acc}=1.0$ to model sequences with $f_\mathrm{acc}=0.6$ with decreasing $\dot{M}$.
\subsubsection{Low accretion rate case} \label{sec:low-rate-case}
\begin{figure}
   \centering
   \small
   \input{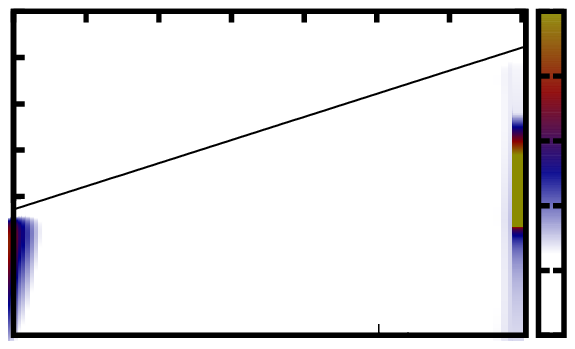}
   \input{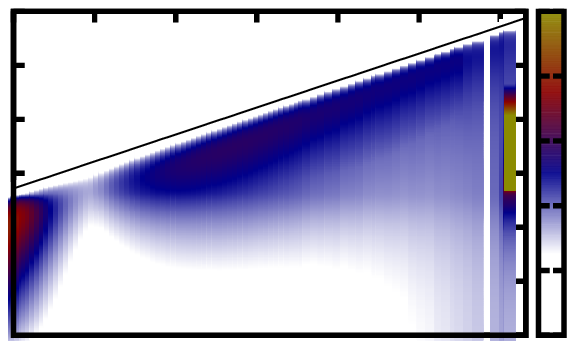}
   \input{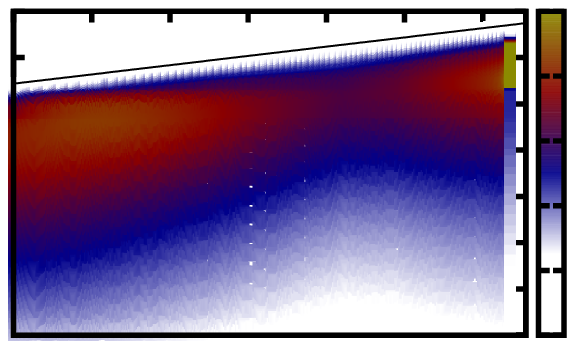}
   \normalsize
      \caption{Temperature profiles over time of individual representative model sequences. $M_\mathrm{r}$ is the mass coordinate, $t$ the time since the start of the accretion phase. The color bar indicates temperature and the black line corresponds to the current total mass of the WD. All model sequences were computed with initial parameters of $M_\mathrm{init}=0.54~\mathrm{M}_\odot$, $\log(L_\mathrm{init}/\mathrm{L}_\odot)=0$ and $f_\mathrm{acc}=1.0$. Model sequence A (L00510-1) uses a mass accretion rate of $\dot{M}=1\cdot10^{-8}~\mathrm{M}_\odot/\mathrm{yr}$, model B (L00510-2) $\dot{M}=2\cdot10^{-8}~\mathrm{M}_\odot/\mathrm{yr}$ and model C (L00510-4) $\dot{M}=4\cdot10^{-8}~\mathrm{M}_\odot/\mathrm{yr}$. It should be noted that the temperature color scale was restricted to the indicated range. Temperatures higher and lower than the limits do occur. The gap visible in plot B is due to a sub sequence of models computed with a larger time step than the rest of the sequence, resulting in a, physically irrelevant, discontinuity of the available data.} 
         \label{fig:tplot}
\end{figure}
\begin{figure}
   \centering
   \small
   \input{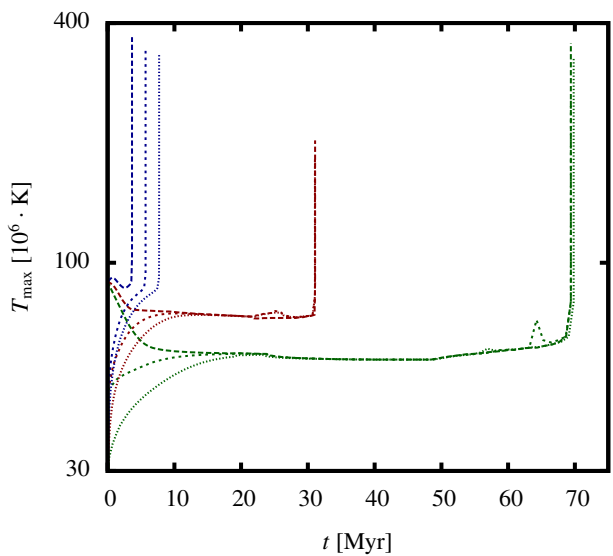}
   \normalsize
      \caption{Maximum temperature over time for a number of representative model sequences. All model depicted model sequences were calculated with $M_\mathrm{init}=0.54~\mathrm{M}_\odot$ and $f_\mathrm{acc}=0.3$. Blue lines indicate a mass accretion rate of $\dot{M}=3\cdot10^{-8}~\mathrm{M}_\odot/\mathrm{yr}$, red lines $\dot{M}=2\cdot10^{-8}~\mathrm{M}_\odot/\mathrm{yr}$ and green lines $\dot{M}=1\cdot10^{-8}~\mathrm{M}_\odot/\mathrm{yr}$. Solid lines indicate an initial luminosity of $\log(L_\mathrm{init}/\mathrm{L}_\odot)=-2$, $\log(L_\mathrm{init}/\mathrm{L}_\odot)=-1$ and dotted lines $\log(L_\mathrm{init}/\mathrm{L}_\odot)=0$.} 
         \label{fig:hplot}
\end{figure} 
As can be seen in Fig.~\ref{fig:fmasses}, systems with $\dot{M}<3 \cdot 10^{-8} \mathrm{M}_\odot/\mathrm{yr}$ exhibit two features absent in the cases with $\dot{M}\geq3 \cdot 10^{-8} \mathrm{M}_\odot/\mathrm{yr}$: \begin{inparaenum} \item The helium shell mass at detonation is much larger than in the corresponding non-rotating case and in the cases with higher mass accretion rates. \item The helium shell mass at detonation is practically independent of the initial luminosity of the WD. \end{inparaenum}

Apart from the already discussed diminished effect of the NCO-reaction, both of these features are explained by the diminished effects of compressional heating at lower mass accretion rates. The compressional heating timescale ($\tau_\mathrm{c}$) is proportional to the mass accretion timescale ($\tau_\mathrm{acc}$), heating being further decreased by rotation \citep[e.g.][]{PSIA2003}. Therefore, if the initial temperature profile of the WD is higher than what can be maintained through the given rate of mass accretion, the WD will cool until an equilibrium between radiative cooling and compressional heating is achieved. If the WD's initial temperature profile is lower than the equilibrium profile, the WD will be heated. Fig.~\ref{fig:tplot} shows that this leads to a period of low temperatures and flat temperature profiles in systems with mass accretion rates $\dot{M}<3 \cdot 10^{-8} \mathrm{M}_\odot/\mathrm{yr}$, while systems with higher mass accretion rates maintain a temperature maximum in the upper parts of their envelope, which in turn aids ignition once an adequate amount of material has been accumulated. This accounts for the significantly increased envelope masses for $\dot{M}<3 \cdot 10^{-8} \mathrm{M}_\odot/\mathrm{yr}$.

Independence from initial luminosity is explained by the same process. Fig.~\ref{fig:hplot} shows the maximum envelope temperature of representative models with respect to mass accretion rates and initial luminosity. As explained above, models with mass accretion rates of $\dot{M}<3 \cdot 10^{-8} \mathrm{M}_\odot/\mathrm{yr}$ evolve towards a state of equilibrium, defined by an equilibrium temperature which depends on the mass accretion rate. Once the equilibrium is reached, further evolution becomes independent of initial parameters. Systems with $\dot{M}\geq3 \cdot 10^{-8} \mathrm{M}_\odot/\mathrm{yr}$ never reach this equilibrium, their evolution being influenced by their initial disposition up to the point of He-ignition.

As will be shown in Sec.~\ref{sec:detonations}, systems conforming to the low accretion rate case are the most likely candidates for He-detonations (as opposed to deflagrations), but their high mass requirements make their natural occurrence highly unlikely. Mass transfer rates of $\dot{M}<3 \cdot 10^{-8} \mathrm{M}_\odot/\mathrm{yr}$ are expected to require a mass donor of $M\leq0.9~\mathrm{M}_\odot$. Such a low-mass donor does not have a mass budget large enough to provide the required amount of helium \citep[see][]{NYL2016}.

\subsubsection{$f_\mathrm{acc}$-dependence} \label{sec:f-dependence}
\begin{figure}
   \centering
   \small
   \input{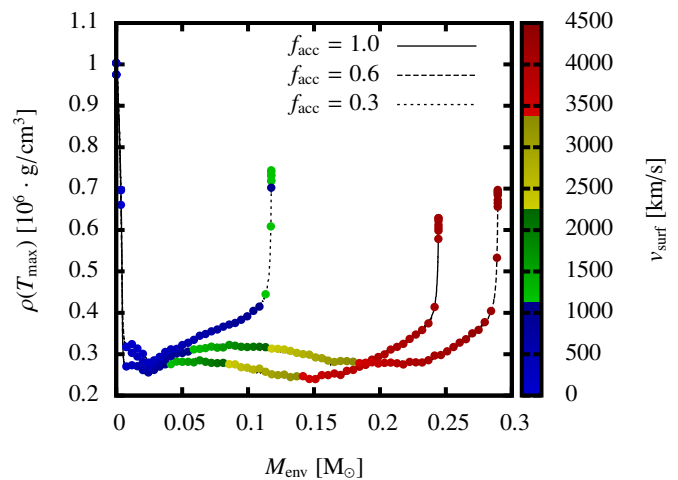}
   \normalsize
      \caption{Densities at the point of highest temperature ($\rho(T_\mathrm{max})$) with respect to the total current mass of the accumulated He-envelope ($M_\mathrm{He}$). The colored dots indicate the current surface velocity of the model. Depicted are model sequences lm10810-3, lm10806-3 and lm10803-3. The model sequences differ in the chosen \mbox{$f_\mathrm{acc}$-parameter}.} 
         \label{fig:rotplotv}
\end{figure} 

\begin{figure}
   \centering
   \small
   \input{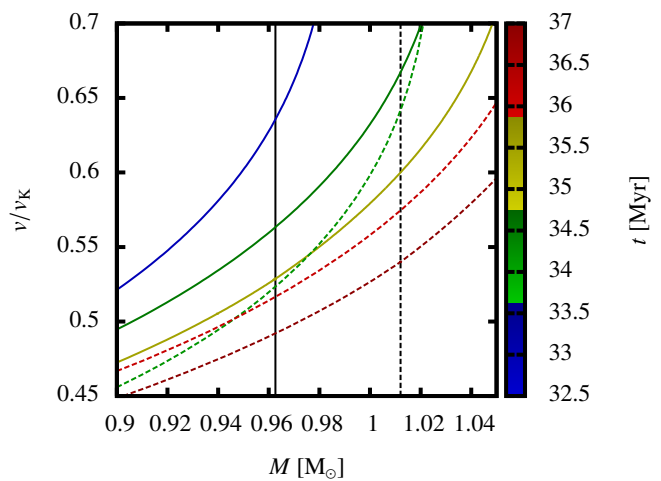}
   \normalsize
      \caption{Local rotational velocity ($v$) divided by the local Keplerian velocity ($v_\mathrm{K}$) with respect to mass coordinate ($M$). Depicted are model sequences lm10810-3 and lm10806-3. Solid lines represent the $f_\mathrm{acc}=1.0$-case (lm10810-3), dotted lines the $f_\mathrm{acc}=0.6$-case (lm10806-3). The color bar indicates the time-coordinate of the profile. The black lines indicate the mass coordinate of the point of ignition for the respective model. The solid black line is valid for $f_\mathrm{acc}=1.0$ and the dotted line for $f_\mathrm{acc}=0.6$.} 
         \label{fig:vvkl}
\end{figure} 

As noted in the previous sections, in the low initial mass and intermediate cases, the ordering with respect to final He-envelope mass of models calculated with different values of the \mbox{$f_\mathrm{acc}$-parameter} depends on the mass accretion rate. In this section we will explain why this switch occurs and why it is related to the mass accretion rate. 

It is important to note that this behavior is unexpected insofar as that, recalling Eq.~\ref{eq:mom-balance}, the effects of rotation should be stronger for more rapid rotation. Since $f_\mathrm{acc}$ controls the amount of angular momentum accumulated per unit mass, a model calculated with a larger value of the \mbox{$f_\mathrm{acc}$-parameter} should be rotating faster than in the case with a smaller $f_\mathrm{acc}$ after the same amount of material has been accreted. This criterion is satisfied in our models. Since the conditions for the onset of helium burning primarily depend on the local density and temperature, a fast, quasi-solidly, rotating star will have to accumulate a larger amount of helium than a slower rotating one in order to facilitate He-ignition. Therefore, naively, one would expect the ignition mass of a given WD to increase with the \mbox{$f_\mathrm{acc}$-parameter}.

Fig.~\ref{fig:rotplotv} shows the evolution of the density at the point of maximum temperature, $\rho(T_\mathrm{max})$, with respect to the mass of the accumulated He-envelope, $M_\mathrm{He}$. Color indicates the surface rotational velocity of the WD (note that the angular velocity, due to quasi solid body rotation, does not differ significantly at the surface and anywhere else in the star). Depicted are three model sequences, all with initial mass $M_\mathrm{init}=0.82~\mathrm{M}_\odot$, initial luminosity $\log(L_\mathrm{init}/\mathrm{L}_\odot)=-1$ and constant mass accretion rate $\dot{M}=3 \cdot 10^{-8} \mathrm{M}_\odot/\mathrm{yr}$ for three different values of $f_\mathrm{acc}$. The initial decrease in density is explained by the fact that, at the start of the accretion phase, the WDs have the highest temperature in the central region. With proceeding accretion of matter, thermal energy is introduced in the outermost layers of the WD while heat is radiated away from the center. This leads to a flattening of the temperature gradient with a peak somewhere in the upper envelope (compare Fig.~\ref{fig:ign06} and Fig.~\ref{fig:ign10}). Once high enough local temperature and density, are reached, the helium will ignite and a flame will propagate inwards, leading to the increase of $\rho(T_\mathrm{max})$ immediately before the end of the model sequence.
Keeping in mind that Eq.~\ref{eq:mom-balance} leads us to expect that slower rotational velocities lead to earlier ignition, the small helium shell mass of the $f_\mathrm{acc}=0.3$ model sequence is easily explained by the observation that this model does not reach rotational velocities close to critical. Both the $f_\mathrm{acc}=1.0$ and the $f_\mathrm{acc}=0.6$ model sequences reach critical rotational velocities, albeit the former sooner than the latter. After the initial decrease, the $f_\mathrm{acc}=1.0$ model sequence exhibits lower values of $\rho(T_\mathrm{max})$ than the $f_\mathrm{acc}=0.6$ model sequence. Due to the relatively higher rotational velocity of the $f_\mathrm{acc}=1.0$ model sequence, this is expected.

The subsequent unexpected increase of $\rho(T_\mathrm{max})$ in the $f_\mathrm{acc}=1.0$ model sequence at an earlier time than in the $f_\mathrm{acc}=0.6$, leading to helium ignition, is because of a combination of the following factors:
As the WD accumulates mass, the radial coordinate of any given mass coordinate decreases, increasing the local Keplerian velocity (Eq.~\ref{eq:kepler}). At the same time, as long as the WD has not reached critical rotation, accumulation of angular momentum through mass accretion increases the local rotational velocity relative to the Keplerian velocity. As discussed in Sec.~\ref{sec:methods}, once a model sequence reaches critical rotation, $f_\mathrm{acc}$ is assumed to be $0$. This means that angular momentum is only accumulated at the rate allowed by the increase of mass of the WD instead of the rate dictated by the \mbox{$f_\mathrm{acc}$-parameter}. This in turn means that, at lower mass coordinates, the local value of $v/v_\mathrm{K}$ decreases due to the increase in value of $v_\mathrm{K}$. This behavior is evident in Fig.~\ref{fig:vvkl}, which shows the evolution of the local value of $v/v_\mathrm{K}$ at the point of ignition with respect to the total mass of the WD for the same model sequences with different values chosen for the \mbox{$f_\mathrm{acc}$-parameter}. The earlier ignition of the model sequence with $f_\mathrm{acc}=1.0$ is therefore precipitated by the fact that it reaches critical rotation earlier than models with a smaller value of $f_\mathrm{acc}$.
\begin{figure}
   \centering
   \small
   \begin{subfigure}[t]{1.0\textwidth}
   \input{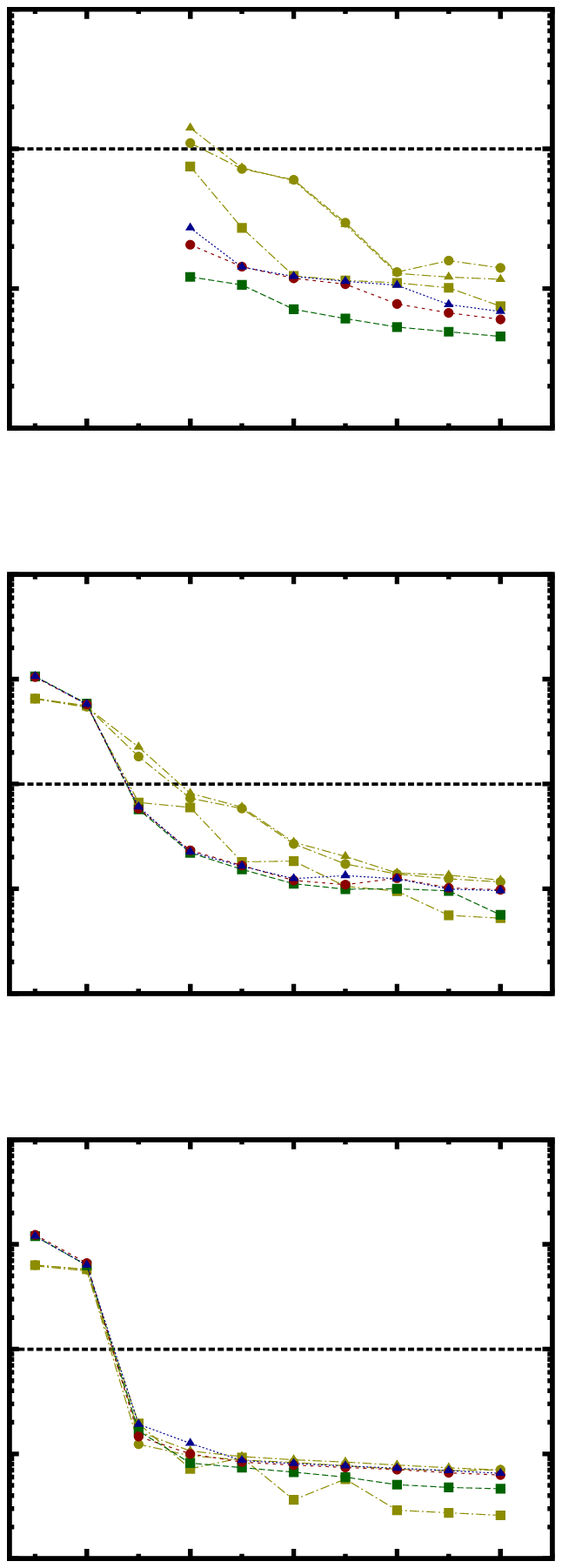}
   \end{subfigure}
   \begin{center}
   \begin{subfigure}[b]{1.0\textwidth}
   \input{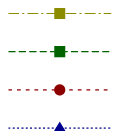}
   \end{subfigure}   
   \end{center}
   \normalsize
      \caption{Density at the point of ignition ($\rho_\mathrm{ign}$) plotted against mass accretion rate. the black dotted line indicates the minimum density for a supersonic runaway according to \cite{WW1994}. Only model sequences with $f_\mathrm{acc}=1.0$ included.}
         \label{fig:rhocritm}
\end{figure}
\subsection{Ignition densities and detonations} \label{sec:detonations}
Fig.~\ref{fig:rhocritm} shows the density at the point of ignition at the time of ignition ($\rho_\mathrm{ign}$) in relation to the chosen mass accretion rate ($\dot{M}$). Most importantly, the expected ignition densities in rotating models are significantly lower than in the non-rotating case. There is a correlation of higher ignition densities with higher initial masses, lower mass accretion rates and lower initial luminosities.

Apart from our baseline non-rotating models, only low-accretion-rate-case systems reach high enough ignition densities for detonation. However, taking into account the fact that the conditions of the mass accretion phase must be present in a physical CO-WD He-Star system in order to produce an observable event, the low-accretion-rate-case presents us with two mutually exclusive conditions for detonation: \begin{inparaenum} \item low accretion rates and \item large amounts ($\sim0.45~\mathrm{M}_\odot$) of transferred mass \end{inparaenum}. As shown by \cite{NYL2016}, mass transfer rates of $\dot{M}\leq3\cdot10^{-8}~\mathrm{M}_\odot/\mathrm{yr}$ are only to expected in close binary systems for He-donors of $M\leq0.9~\mathrm{M}_\odot$, varying with the evolution of the donor. Such a donor star would need to lose a majority of its mass in order to produce a detonation. The same criterion rules out the possibility that the required low mass accretion rates could be provided through wind accretion from a more massive donor. Previous studies \citep[e.g.][]{TBA1996} suggest that a companion star might be able to accrete a few percent ($\sim1\dots2\%$) of a primary's wind mass loss rate. A He-star exhibiting sufficiently high wind mass loss to lead to an accretion rate of $\sim1\cdot10^{-8}~\mathrm{M}_\odot/\mathrm{yr}$ on an accompanying WD would be unlikely to be able to sustain it for long enough to allow for the accumulation of $\sim0.5~\mathrm{M}_\odot$ of He.

We note that our non-rotating models, including the predictions of whether the WD is capable of producing a detonation, are in reasonably good agreement with previous results such as those obtained by \cite{WK2011}, generally showing a discrepancy of less than 5\% in final helium shell mass. Larger discrepancies of up to 30\% in final helium shell mass in the comparable model sequences of initial masses around $0.8~\mathrm{M}_\odot$ at mass accretion rates of $2 \dots 3\cdot10^{-8}\mathrm{M}/\mathrm{yr}$ can be attributed to the presence of more massive He-layer in our $0.82~\mathrm{M}_\odot$-model. Comparison with comparable models obtained by \cite{PTY2014} yield a discrepancy of $5 \dots 20$\%, which can be attributed to the higher initial luminosities used in that source. We also note that the chosen methodology does not take into account changes in the mass accretion rate, which would occur in any natural interacting binary. We nevertheless conclude that, while it may be theoretically possible to induce a He-detonation on a rotating WD with an active Tayler-Spruit dynamo, the conditions necessary to do so are unlikely to exist in a naturally occurring binary system. Any naturally occurring accretor will far more likely undergo a helium deflagration as opposed to a detonation. However, we stress again the limitations of using the prescription introduced in \cite{WW1994} and point to \cite{WK2011} where detonations have been found to occur at lower (albeit not much lower) ignition densities than the $\rho_\mathrm{crit}=10^{6}~\mathrm{g}/\mathrm{cm}^{3}$ used here. More detailed calculations using a more specialized simulation framework would have to be used in order to resolve this problem. 

As far as resulting transients are concerned, we find it unlikely that any of the rotating models under consideration here might be able to produce classical Type\,Ia SNe. Returning to \cite{WK2011} for reference (although due to the reference only discussing non-rotating WDs, this comparison must be considered rough), Type Ia-like spectra and light curves depend on low He-shell masses $\lesssim0.05~\mathrm{M}_\odot$ at ignition as well as the ignition turning into a supersonic shock. As discussed above, comparable He-shell masses only occur in our model sequences with $M_\mathrm{init}\geq1.0~\mathrm{M}_\odot$ and $\dot{M}\geq5\cdot10^{-8}~\mathrm{M}_\odot/\mathrm{yr}$, which are incapable of producing a detonation when rotation is included. On the other hand, the only systems in our sample which are able to produce a detonation are systems with $M_\mathrm{init}\leq0.82~\mathrm{M}_\odot$ at mass accretion rates $\dot{M}\leq3 \cdot 10^{-8} \mathrm{M}_\odot/\mathrm{yr}$. But the He-shell masses at detonation of these systems --- which, as explained above, we do not expect to occur naturally anyhow --- are large enough to render a SN Ia-like spectrum very unlikely.

We emphasize at this point that our parameter space excludes accretion rates of $\dot{M}<4\cdot10^{-8}~\mathrm{M}_\odot/\mathrm{yr}$ for WDs with $M_\mathrm{init}\geq1.0~\mathrm{M}_\odot$. Although we cannot specifically exclude ignitions developing into detonations in this part of the parameter space, the resulting He-shell masses would nevertheless suggest resulting spectra unlike those of classical SNe Ia.

However, the comparatively high ignition masses and ignition densities just below $10^6\,$g/cm$^3$, in systems with accretion rates of $\dot{M}\simeq 2\dots 3 \cdot 10^{-8} \mathrm{M}_\odot/\mathrm{yr}$, could correspond acceptably well with the properties of the fast, faint calcium-rich SNe, as discussed by \cite{WSL2011}.

\section{Comparison with the non-magnetic rotating case} \label{sec:comp}
As mentioned before, the impact of rotation with non-magnetic angular momentum diffusion and energy dissipation, the non-magnetic rotating case, has been studied by \cite{YL2004b}, which differs from the systems in this study, the magnetic rotating case, in its neglect of the Tayler-Spruit mechanism.
 
\begin{table} [h]
\centering
\caption{Comparison between model sequences obtained by \cite{YL2004b} (A) and this~study~(B):\\
Initial mass $M_\mathrm{WD,i}$, mass accretion rate $\dot{M}$, $f_\mathrm{acc}$-parameter, Helium shell mass at ignition $M_\mathrm{He,f}$ and whether the system is expected to detonate. \\
All systems were calculated with an initial luminosity of $\log{L_\mathrm{init}/\mathrm{L}_\odot}=-2$. }
\label{tab:compare}
\begin{tabular}{l l l l l l}
\hline\hline
Study & $M_\mathrm{WD,i}$ & $\dot{M}$                                & $f_\mathrm{acc}$ & $M_\mathrm{He,f}$ & Det? \\
      & $M_\odot$       & $[10^{-8}\cdot\mathrm{M}_\odot/\mathrm{yr}]$ &                & $M_\odot$       &             \\
\hline 
A & 0.60 & 2 & 1.0 & 0.08 & No \\
A & 0.60 & 3 & 1.0 & 0.06 & No \\
A & 0.80 & 2 & 1.0 & 0.02 & No \\
A & 0.80 & 3 & 1.0 & 0.02 & No \\
\hline
A & 0.60 & 2 & 0.6 & 0.15 & No \\
A & 0.60 & 3 & 0.6 & 0.11 & No \\
A & 0.80 & 2 & 0.6 & 0.07 & No \\
A & 0.80 & 3 & 0.6 & 0.05 & No \\
\hline
\hline
B & 0.54 & 2 & 1.0 & 0.61 & Yes \\
B & 0.54 & 3 & 1.0 & 0.23 & No \\
B & 0.82 & 2 & 1.0 & 0.42 & Yes \\
B & 0.82 & 3 & 1.0 & 0.25 & No \\
\hline
B & 0.54 & 2 & 0.6 & 0.62 & Yes \\
B & 0.54 & 3 & 0.6 & 0.24 & No \\
B & 0.82 & 2 & 0.6 & 0.42 & No \\
B & 0.82 & 3 & 0.6 & 0.32 & No \\
\hline
\end{tabular}
\end{table}
Tab.~\ref{tab:compare} allows a comparison between systems taken from \cite{YL2004b} and this study. While there are some differences in the initial mass of the model sequences, these are sufficiently small to make this comparison meaningful.

Comparing models with $f_\mathrm{acc}=1.0$, it is evident that the amount of helium required to induce ignition is significantly smaller in the rotating, non-magnetic case than in the rotating, magnetic case. The impact of the inclusion of the Tayler-Spruit mechanism is drastic, with the state of initially very similar systems being very different at the point of helium ignition. For example, unlike in the rotating, non-magnetic case, where sufficiently high ignition densities are precluded by a strong, localized input of thermal energy 
through energy dissipation near the core-shell interface (see Sec.~\ref{sec:ignitions}), given a sufficiently large mass budget, a rotating, magnetic system would actually result in a helium detonation.

Another important difference is that in the rotating, non-magnetic case, the amount of accreted helium until ignition is smaller when compared to the non-rotating case, while it is larger in the rotating, magnetic case.

However, in both the rotating, non-magnetic case and the rotating, magnetic case, we expect He-detonations to be an unlikely outcome of the WD's evolution. In the rotating, non-magnetic case this is a consequence of the already mentioned localized input of thermal energy through viscous dissipation. In the rotating, magnetic case, the reason lies in the flatter density stratification brought about by quasi-solid-body rotation in cases with $\dot{M}\geq3 \cdot 10^{-8} \mathrm{M}_\odot/\mathrm{yr}$ and the expected insufficient mass budget of prospective mass donors in cases with $\dot{M}\leq3 \cdot 10^{-8} \mathrm{M}_\odot/\mathrm{yr}$. 
Due to the significant differences in the He-shell mass at ignition, the transients resulting from helium ignition in rotating, non-megnatic and rotating-magnetic systems would differ significantly from each other.

\section{Evolutionary fates} \label{sec:outcomes}
As described in Sec.~\ref{sec:detonations}, our magnetic, rotating WDs are not promising progenitor models for classical Type\,Ia supernovae, or for other potential transient requiring the ignition of the underlying CO core. The question naturally arises: What do they result in?

To discuss this question, we may divide the parameter space of our models into three parts: the low accretion rate regime, i.e., $\dot{M}\leq 2\cdot 10^{-8} \mathrm{M}_\odot/\mathrm{yr}$, the high accretion rate regime, i.e., $\dot{M}\geq 4\cdot 10^{-8} \mathrm{M}_\odot/\mathrm{yr}$, and the transition regime in between the first two.

In the low accretion rate regime, ignition densities would be sufficiently high to produce detonations. However, as argued above, since for the ignition to occur, a high mass of helium needs to be accreted, this may not happen in nature. Instead, the accretion may stop before the helium can ignite \citep{NYL2016}, leading to a close binary consisting of a CO white dwarf covered by a degenerate helium envelope and a He white dwarf with some amounts of CO in its core. Gravitational wave radiation may lead to a merger of such binaries, which could either produce a cataclysmic event \citep[e.g.][]{KCJ2010,DRB2014} or an R\,CrB star \citep[e.g.][]{SMH2012, MHD2013}.

In the high accretion rate regime, our rotating models accrete only  $0.01 \dots 0.2\,$M$_{\odot}$ until helium ignites (Fig.\,7), which occurs at densities below $\sim 3\cdot10^5\,$g/cm$^3$ (Fig.\,12) and renders helium detonations unlikely. Such events might be comparable to the faint He-flashes discussed by \cite{BSWN2007} in the context of AM CVn systems. 

In the transition regime, however, we find some rotating models which ignite helium after only accreting slightly more than $0.2\,$M$_{\odot}$, at a density above $\sim 5\cdot10^5\,$g/cm$^3$ (e.g., Models Lm20810-3, Lm10810-3, L00810-3). When our density criterion for He detonation is applied to these models, a helium shell deflagration is predicted to occur. While our density criterion is uncertain (Sect.\,2.3), we can not exclude that a detonation would occur in this case. However, the ignition conditions in these models are similar to those investigated by \cite{WSL2011}, who argue that even in case a helium detonation would occur, the CO core might remain unaffected in this situation.

It was shown by \cite{WSL2011} that the helium shell explosion with parameters corresponding to our transition regime models, even if it occurs as a detonation, leads to an incomplete burning of the accumulated helium, with a rich production of intermediate mass elements \citep[see also][]{HHS1983}. \cite{WSL2011} propagate that the resulting explosions may correspond to the observationally identified class of Ca-rich supernovae, which are a fast and faint sub-type of Type\,Ib supernovae \citep{PGM2010}, although arguments favoring a core-collapse origin have been put forward as well \citep{KMN2010}.

Overall, it appears unlikely in the light of our models that helium accreting white dwarfs provide the conditions for classical Type\,Ia supernovae. Instead, they may be good progenitors for fast and faint hydrogen-free transient events, with helium novae occuring for the lowest accumulated helium masses ($\leq 0.01\,$M$_{\odot}$) at the faint end, up to sub-luminous Type\,I supernovae like the Ca-rich supernovae at the bright end. 

It has been shown observationally by \cite{PGC2011} that Ca-rich supernovae tend to be offset from their host galaxies by a considerable distance and located in regions where low-mass progenitors in old stellar populations would be expected. The corresponding progenitor age is subject to an ongoing debate. \cite{YKS2013} argue for delay times of $10~\mathrm{Gyr}$ or more, while \cite{SYJY2011} give characteristic ages as low as $300~\mathrm{Myr}$. Calculating delay times for the systems under consideration here is beyond our scope, but \cite{WJH2013} used similar masses in their study, but neglecting rotation. According to their estimates, a delay time for sub-Chandrasekhar mass detonations in He+WD systems close to $1~\mathrm{Gyr}$ is most probable. Questions concerning the delay times of these transients are equally applicable to white dwarf as to core-collapse progenitors \citep{YKS2013,ZMI2017} and will have to be addressed in future research.

\section{Conclusions} \label{sec:conclusions}
We have carried out detailed simulations of CO-WDs accreting helium-rich matter at constant accretion rates 
of $10^{-8}\dots 10^{-7} \mathrm{M}_\odot/\mathrm{yr}$, allowing for the accumulation of angular momentum and rotational energy dissipation and considering the effects of the Tayler-Spruit mechanism, dynamical shear instability, secular shear instability and the Goldreich-Schubert-Fricke instability up to the point of He-ignition.

We find that inclusion of the Tayler-Spruit mechanism profoundly impacts the expected rotational profile of the WD, inducing quasi-solid body rotation on the object. We also find that, due to the apparent rotational profile, the role of the NCO-effect in facilitating ignition of the helium envelope is diminished compared to the non-rotating case. Our calculations show a stark difference in the relation of the amount of helium accreted at the point of ignition between rotating and non-rotating model sequences. The magnitude of the difference depends on the initial mass of the WD. Low mass WDs generally accrete up to $0.1~\mathrm{M}_\odot$ more ($0.3~\mathrm{M}_\odot$ in total) at accretion rates of $\dot{M}\geq3 \cdot 10^{-8} \mathrm{M}_\odot/\mathrm{yr}$ and up to $0.2~\mathrm{M}_\odot$ more ($0.7~\mathrm{M}_\odot$ in total) at accretion rates of $\dot{M}<3 \cdot 10^{-8} \mathrm{M}_\odot/\mathrm{yr}$. High mass WDs accrete on the order of $0.01~\mathrm{M}_\odot$ less than in the non-rotating case.

Rotating, magnetic systems are also found to exhibit significantly increased ignition masses when compared to previous non-magnetic rotating models \citep{YL2004b}.

We find that rotating, magnetic WDs, accreting helium at rates $\dot{M}\geq3 \cdot 10^{-8} \mathrm{M}_\odot/\mathrm{yr}$, will not reach ignition densities above $10^6~\mathrm{g}/\mathrm{cm}^3$, which indicates that helium ignitions in these systems are unlikely to develop into a detonation. Only systems with $\dot{M}\leq2 \cdot 10^{-8} \mathrm{M}_\odot/\mathrm{yr}$ exhibit ignition densities larger than $10^6~\mathrm{g}/\mathrm{cm}^3$. However, naturally occurring systems of this type are unlikely to eventually produce a He-detonation since the expected large amount ($\gtrsim0.5~\mathrm{M}_\odot$) of helium required to induce helium ignition would call for a suitably massive donor star. Such a donor would in turn be unable to provide sufficiently low mass transfer rates \citep{NYL2016}.

This means that identifying any of these models as the progenitors of classical SNe Ia via the double detonation mechanism is not seen as viable. However, the obtained ignition masses and densities for rotating magnetic WDs with 
accretion rates of $\dot{M}\simeq 3 \cdot 10^{-8} \mathrm{M}_\odot/\mathrm{yr}$ resemble the conditions which appear promising to reproduce the faint and fast Ca-rich supernovae. More research, especially concerning explosion physics, would be desirable in order to strengthen this indication.

In any case, the stark difference between the expected states of the accreting WD at the time of ignition between the the rotating, magnetic case, as discussed in this study, the rotating, non-magnetic case \citep[e.g.][]{YL2004b} and the non-rotating case \citep[e.g.][]{WK2011} are a clear indication that rotation and magnetic effects should not be neglected in the future study of accretion scenarios in the lead-up to explosive stellar transients.

\begin{acknowledgements}
This research was supported by the Korea Astronomy and Space Science Institute
under the R\&D program (Project No. 3348-20160002) supervised by the Ministry
of Science, ICT and Future Planning.
Support by the Deutsche Forschungsgemeinschaft (DFG), Grant No. Yo 194/1-1, is gratefully acknowledged.
The authors wish to thank the anonymous referee, whose comments added to the clarity and scope of the article.
PN would like to thank Phillip Podsiadlowski for useful discussions.
\end{acknowledgements}

\bibliographystyle{aa}
\bibliography{paper.bib}{}

\begin{thebibliography}{88}
\expandafter\ifx\csname natexlab\endcsname\relax\def\natexlab#1{#1}\fi

\bibitem[{{Benz}(1997)}]{B1997P}
{Benz}, W. 1997, in NATO Advanced Science Institutes (ASI) Series C, Vol. 486,
  NATO Advanced Science Institutes (ASI) Series C, ed. P.~{Ruiz-Lapuente},
  R.~{Canal}, \& J.~{Isern}, 457

\bibitem[{{Bildsten} {et~al.}(2007){Bildsten}, {Shen}, {Weinberg}, \&
  {Nelemans}}]{BSWN2007}
{Bildsten}, L., {Shen}, K.~J., {Weinberg}, N.~N., \& {Nelemans}, G. 2007,
  \apjl, 662, L95

\bibitem[{Brooks {et~al.}(2016)Brooks, Bildsten, Schwab, \& Paxton}]{BBS2016}
Brooks, J., Bildsten, L., Schwab, J., \& Paxton, B. 2016, The Astrophysical
  Journal, 821, 28

\bibitem[{{Dan} {et~al.}(2014){Dan}, {Rosswog}, {Br{\"u}ggen}, \&
  {Podsiadlowski}}]{DRB2014}
{Dan}, M., {Rosswog}, S., {Br{\"u}ggen}, M., \& {Podsiadlowski}, P. 2014,
  \mnras, 438, 14

\bibitem[{{Eggenberger} {et~al.}(2005){Eggenberger}, {Maeder}, \&
  {Meynet}}]{EMM2005}
{Eggenberger}, P., {Maeder}, A., \& {Meynet}, G. 2005, \aap, 440, L9

\bibitem[{{Endal} \& {Sofia}(1976)}]{ES1976}
{Endal}, A.~S. \& {Sofia}, S. 1976, \apj, 210, 184

\bibitem[{{Endal} \& {Sofia}(1978)}]{ES1978}
{Endal}, A.~S. \& {Sofia}, S. 1978, \apj, 220, 279

\bibitem[{{Fink} {et~al.}(2007){Fink}, {Hillebrandt}, \& {R{\"o}pke}}]{FHR2007}
{Fink}, M., {Hillebrandt}, W., \& {R{\"o}pke}, F.~K. 2007, \aap, 476, 1133

\bibitem[{{Fink} {et~al.}(2010){Fink}, {R{\"o}pke}, {Hillebrandt},
  {Seitenzahl}, {Sim}, \& {Kromer}}]{FRH2010}
{Fink}, M., {R{\"o}pke}, F.~K., {Hillebrandt}, W., {et~al.} 2010, \aap, 514,
  A53

\bibitem[{{Foley} {et~al.}(2013){Foley}, {Challis}, {Chornock},
  {Ganeshalingam}, {Li}, {Marion}, {Morrell}, {Pignata}, {Stritzinger},
  {Silverman}, {Wang}, {Anderson}, {Filippenko}, {Freedman}, {Hamuy}, {Jha},
  {Kirshner}, {McCully}, {Persson}, {Phillips}, {Reichart}, \&
  {Soderberg}}]{FCC2013}
{Foley}, R.~J., {Challis}, P.~J., {Chornock}, R., {et~al.} 2013, \apj, 767, 57

\bibitem[{{Foley} {et~al.}(2016){Foley}, {Jha}, {Pan}, {Zheng}, {Bildsten},
  {Filippenko}, \& {Kasen}}]{FJP2016}
{Foley}, R.~J., {Jha}, S.~W., {Pan}, Y.-C., {et~al.} 2016, \mnras, 461, 433

\bibitem[{{Foley} {et~al.}(2015){Foley}, {Van Dyk}, {Jha}, {Clubb},
  {Filippenko}, {Mauerhan}, {Miller}, \& {Smith}}]{FDJ2015}
{Foley}, R.~J., {Van Dyk}, S.~D., {Jha}, S.~W., {et~al.} 2015, \apjl, 798, L37

\bibitem[{{Hashimoto} {et~al.}(1983){Hashimoto}, {Hanawa}, \&
  {Sugimoto}}]{HHS1983}
{Hashimoto}, M.-A., {Hanawa}, T., \& {Sugimoto}, D. 1983, \pasj, 35, 1

\bibitem[{{Hashimoto} {et~al.}(1986){Hashimoto}, {Nomoto}, {Arai}, \&
  {Kaminisi}}]{HNAK1986}
{Hashimoto}, M.-A., {Nomoto}, K.-I., {Arai}, K., \& {Kaminisi}, K. 1986, \apj,
  307, 687

\bibitem[{{Heber} {et~al.}(1997){Heber}, {Napiwotzki}, \& {Reid}}]{HNR1997}
{Heber}, U., {Napiwotzki}, R., \& {Reid}, I.~N. 1997, \aap, 323, 819

\bibitem[{{Heger} \& {Langer}(2000)}]{HL2000}
{Heger}, A. \& {Langer}, N. 2000, \apj, 544, 1016

\bibitem[{{Heger} {et~al.}(2000){Heger}, {Langer}, \& {Woosley}}]{HLW2000}
{Heger}, A., {Langer}, N., \& {Woosley}, S.~E. 2000, \apj, 528, 368

\bibitem[{{Heger} {et~al.}(2005){Heger}, {Woosley}, \& {Spruit}}]{HWS2005}
{Heger}, A., {Woosley}, S.~E., \& {Spruit}, H.~C. 2005, The Astrophysical
  Journal, 626, 350

\bibitem[{{Ivanova} {et~al.}(2013){Ivanova}, {Justham}, {Chen}, {De Marco},
  {Fryer}, {Gaburov}, {Ge}, {Glebbeek}, {Han}, {Li}, {Lu}, {Marsh},
  {Podsiadlowski}, {Potter}, {Soker}, {Taam}, {Tauris}, {van den Heuvel}, \&
  {Webbink}}]{IJC2013}
{Ivanova}, N., {Justham}, S., {Chen}, X., {et~al.} 2013, \aapr, 21, 59

\bibitem[{Kato \& Hachisu(2004)}]{HK2004}
Kato, M. \& Hachisu, I. 2004, The Astrophysical Journal Letters, 613, L129

\bibitem[{{Kawabata} {et~al.}(2010){Kawabata}, {Maeda}, {Nomoto},
  {Taubenberger}, {Tanaka}, {Deng}, {Pian}, {Hattori}, \& {Itagaki}}]{KMN2010}
{Kawabata}, K.~S., {Maeda}, K., {Nomoto}, K., {et~al.} 2010, \nat, 465, 326

\bibitem[{{Kawaler}(2003)}]{K2003}
{Kawaler}, S.~D. 2003, ArXiv Astrophysics e-prints

\bibitem[{{Kelly} {et~al.}(2014){Kelly}, {Fox}, {Filippenko}, {Cenko}, {Prato},
  {Schaefer}, {Shen}, {Zheng}, {Graham}, \& {Tucker}}]{KFF2014}
{Kelly}, P.~L., {Fox}, O.~D., {Filippenko}, A.~V., {et~al.} 2014, \apj, 790, 3

\bibitem[{{Kilic} {et~al.}(2014){Kilic}, {Hermes}, {Gianninas}, {Brown},
  {Heinke}, {Ag{\"u}eros}, {Chote}, {Sullivan}, {Bell}, \& {Harrold}}]{KHG2014}
{Kilic}, M., {Hermes}, J.~J., {Gianninas}, A., {et~al.} 2014, \mnras, 438, L26

\bibitem[{{Kippenhahn} \& {Thomas}(1978)}]{KT1978}
{Kippenhahn}, R. \& {Thomas}, H.-C. 1978, \aap, 63, 265

\bibitem[{{Koester} {et~al.}(1998){Koester}, {Dreizler}, {Weidemann}, \&
  {Allard}}]{KDW1998}
{Koester}, D., {Dreizler}, S., {Weidemann}, V., \& {Allard}, N.~F. 1998, \aap,
  338, 612

\bibitem[{{Kromer} {et~al.}(2015){Kromer}, {Ohlmann}, {Pakmor}, {Ruiter},
  {Hillebrandt}, {Marquardt}, {R{\"o}pke}, {Seitenzahl}, {Sim}, \&
  {Taubenberger}}]{KOP2015}
{Kromer}, M., {Ohlmann}, S.~T., {Pakmor}, R., {et~al.} 2015, \mnras, 450, 3045

\bibitem[{{Kromer} {et~al.}(2010){Kromer}, {Sim}, {Fink}, {R{\"o}pke},
  {Seitenzahl}, \& {Hillebrandt}}]{KSF2010}
{Kromer}, M., {Sim}, S.~A., {Fink}, M., {et~al.} 2010, \apj, 719, 1067

\bibitem[{{Langer} {et~al.}(2000){Langer}, {Deutschmann}, {Wellstein}, \&
  {H{\"o}flich}}]{LDWH2000}
{Langer}, N., {Deutschmann}, A., {Wellstein}, S., \& {H{\"o}flich}, P. 2000,
  \aap, 362, 1046

\bibitem[{{Li} {et~al.}(2011){Li}, {Bloom}, {Podsiadlowski}, {Miller}, {Cenko},
  {Jha}, {Sullivan}, {Howell}, {Nugent}, {Butler}, {Ofek}, {Kasliwal},
  {Richards}, {Stockton}, {Shih}, {Bildsten}, {Shara}, {Bibby}, {Filippenko},
  {Ganeshalingam}, {Silverman}, {Kulkarni}, {Law}, {Poznanski}, {Quimby},
  {McCully}, {Patel}, {Maguire}, \& {Shen}}]{LBP2011}
{Li}, W., {Bloom}, J.~S., {Podsiadlowski}, P., {et~al.} 2011, \nat, 480, 348

\bibitem[{{Li} {et~al.}(2003){Li}, {Filippenko}, {Chornock}, {Berger},
  {Berlind}, {Calkins}, {Challis}, {Fassnacht}, {Jha}, {Kirshner}, {Matheson},
  {Sargent}, {Simcoe}, {Smith}, \& {Squires}}]{LFA2003}
{Li}, W., {Filippenko}, A.~V., {Chornock}, R., {et~al.} 2003, \pasp, 115, 453

\bibitem[{{Livne}(1990)}]{L1990}
{Livne}, E. 1990, \apjl, 354, L53

\bibitem[{{Livne}(1997)}]{L1997P}
{Livne}, E. 1997, in NATO Advanced Science Institutes (ASI) Series C, Vol. 486,
  NATO Advanced Science Institutes (ASI) Series C, ed. P.~{Ruiz-Lapuente},
  R.~{Canal}, \& J.~{Isern}, 425

\bibitem[{{Livne} \& {Arnett}(1995)}]{LA1995}
{Livne}, E. \& {Arnett}, D. 1995, \apj, 452, 62

\bibitem[{{Livne} \& {Glasner}(1990)}]{LG1990}
{Livne}, E. \& {Glasner}, A.~S. 1990, \apj, 361, 244

\bibitem[{{Livne} \& {Glasner}(1991)}]{LG1991}
{Livne}, E. \& {Glasner}, A.~S. 1991, \apj, 370, 272

\bibitem[{{Maeder} \& {Meynet}(2003)}]{MM2003}
{Maeder}, A. \& {Meynet}, G. 2003, \aap, 411, 543

\bibitem[{{Maeder} \& {Meynet}(2004)}]{MM2004}
{Maeder}, A. \& {Meynet}, G. 2004, \aap, 422, 225

\bibitem[{{Magee} {et~al.}(2016){Magee}, {Kotak}, {Sim}, {Kromer},
  {Rabinowitz}, {Smartt}, {Baltay}, {Campbell}, {Chen}, {Fink}, {Gal-Yam},
  {Galbany}, {Hillebrandt}, {Inserra}, {Kankare}, {Le Guillou}, {Lyman},
  {Maguire}, {Pakmor}, {R{\"o}pke}, {Ruiter}, {Seitenzahl}, {Sullivan},
  {Valenti}, \& {Young}}]{MKS2016}
{Magee}, M.~R., {Kotak}, R., {Sim}, S.~A., {et~al.} 2016, \aap, 589, A89

\bibitem[{{Maoz} \& {Mannucci}(2008)}]{MM2008}
{Maoz}, D. \& {Mannucci}, F. 2008, \mnras, 388, 421

\bibitem[{{McCully} {et~al.}(2014){McCully}, {Jha}, {Foley}, {Bildsten},
  {Fong}, {Kirshner}, {Marion}, {Riess}, \& {Stritzinger}}]{MJS2014}
{McCully}, C., {Jha}, S.~W., {Foley}, R.~J., {et~al.} 2014, \nat, 512, 54

\bibitem[{{Menon} {et~al.}(2013){Menon}, {Herwig}, {Denissenkov}, {Clayton},
  {Staff}, {Pignatari}, \& {Paxton}}]{MHD2013}
{Menon}, A., {Herwig}, F., {Denissenkov}, P.~A., {et~al.} 2013, \apj, 772, 59

\bibitem[{{Mochkovitch} \& {Livio}(1989)}]{ML1989}
{Mochkovitch}, R. \& {Livio}, M. 1989, \aap, 209, 111

\bibitem[{{Mosser} {et~al.}(2012){Mosser}, {Goupil}, {Belkacem}, {Marques},
  {Beck}, {Bloemen}, {De Ridder}, {Barban}, {Deheuvels}, {Elsworth}, {Hekker},
  {Kallinger}, {Ouazzani}, {Pinsonneault}, {Samadi}, {Stello}, {García},
  {Klaus}, {Li}, {Mathur}, \& {Morris}}]{MGB2012}
{Mosser}, B., {Goupil}, M.~J., {Belkacem}, K., {et~al.} 2012, A\&A, 548, A10

\bibitem[{{Nandkumar} \& {Pethick}(1984)}]{NP1984}
{Nandkumar}, R. \& {Pethick}, C.~J. 1984, \mnras, 209, 511

\bibitem[{{Nelemans}(2005)}]{N2005}
{Nelemans}, G. 2005, in Astronomical Society of the Pacific Conference Series,
  Vol. 330, The Astrophysics of Cataclysmic Variables and Related Objects, ed.
  J.-M. {Hameury} \& J.-P. {Lasota}, 27

\bibitem[{{Nelemans} {et~al.}(2008){Nelemans}, {Voss}, {Roelofs}, \&
  {Bassa}}]{NVR2008}
{Nelemans}, G., {Voss}, R., {Roelofs}, G., \& {Bassa}, C. 2008, \mnras, 388,
  487

\bibitem[{{Neunteufel} {et~al.}(2016){Neunteufel}, {Yoon}, \&
  {Langer}}]{NYL2016}
{Neunteufel}, P., {Yoon}, S.-C., \& {Langer}, N. 2016, \aap, 589, A43

\bibitem[{{Nomoto}(1980)}]{N1980P}
{Nomoto}, K. 1980, in Texas Workshop on Type I Supernovae, ed. J.~C. {Wheeler},
  164--181

\bibitem[{{Nomoto}(1982{\natexlab{a}})}]{N1982b}
{Nomoto}, K. 1982{\natexlab{a}}, \apj, 257, 780

\bibitem[{{Nomoto}(1982{\natexlab{b}})}]{N1982a}
{Nomoto}, K. 1982{\natexlab{b}}, \apj, 253, 798

\bibitem[{{Paczynski}(1991)}]{P1991}
{Paczynski}, B. 1991, \apj, 370, 597

\bibitem[{{Perets} {et~al.}(2011){Perets}, {Gal-yam}, {Crockett}, {Anderson},
  {James}, {Sullivan}, {Neill}, \& {Leonard}}]{PGC2011}
{Perets}, H.~B., {Gal-yam}, A., {Crockett}, R.~M., {et~al.} 2011, \apjl, 728,
  L36

\bibitem[{{Perets} {et~al.}(2010){Perets}, {Gal-Yam}, {Mazzali}, {Arnett},
  {Kagan}, {Filippenko}, {Li}, {Arcavi}, {Cenko}, {Fox}, {Leonard}, {Moon},
  {Sand}, {Soderberg}, {Anderson}, {James}, {Foley}, {Ganeshalingam}, {Ofek},
  {Bildsten}, {Nelemans}, {Shen}, {Weinberg}, {Metzger}, {Piro}, {Quataert},
  {Kiewe}, \& {Poznanski}}]{PGM2010}
{Perets}, H.~B., {Gal-Yam}, A., {Mazzali}, P.~A., {et~al.} 2010, \nat, 465, 322

\bibitem[{{Perlmutter} {et~al.}(1999){Perlmutter}, {Aldering}, {Goldhaber},
  {Knop}, {Nugent}, {Castro}, {Deustua}, {Fabbro}, {Goobar}, {Groom}, {Hook},
  {Kim}, {Kim}, {Lee}, {Nunes}, {Pain}, {Pennypacker}, {Quimby}, {Lidman},
  {Ellis}, {Irwin}, {McMahon}, {Ruiz-Lapuente}, {Walton}, {Schaefer}, {Boyle},
  {Filippenko}, {Matheson}, {Fruchter}, {Panagia}, {Newberg}, {Couch}, \&
  {Project}}]{PAG1999}
{Perlmutter}, S., {Aldering}, G., {Goldhaber}, G., {et~al.} 1999, \apj, 517,
  565

\bibitem[{{Piersanti} {et~al.}(2003){Piersanti}, {Gagliardi}, {Iben}, \&
  {Tornamb{\'e}}}]{PSIA2003}
{Piersanti}, L., {Gagliardi}, S., {Iben}, Jr., I., \& {Tornamb{\'e}}, A. 2003,
  \apj, 583, 885

\bibitem[{{Piersanti} {et~al.}(2014){Piersanti}, {Tornamb{\'e}}, \&
  {Yungelson}}]{PTY2014}
{Piersanti}, L., {Tornamb{\'e}}, A., \& {Yungelson}, L.~R. 2014, \mnras, 445,
  3239

\bibitem[{{Piersanti} {et~al.}(2015){Piersanti}, {Yungelson}, \&
  {Tornamb{\'e}}}]{PYT2015}
{Piersanti}, L., {Yungelson}, L.~R., \& {Tornamb{\'e}}, A. 2015, \mnras, 452,
  2897

\bibitem[{{Popham} \& {Narayan}(1991)}]{PN1991}
{Popham}, R. \& {Narayan}, R. 1991, \apj, 370, 604

\bibitem[{{Renedo} {et~al.}(2010){Renedo}, {Althaus}, {Miller Bertolami},
  {Romero}, {C{\'o}rsico}, {Rohrmann}, \& {Garc{\'{\i}}a-Berro}}]{RAB2010}
{Renedo}, I., {Althaus}, L.~G., {Miller Bertolami}, M.~M., {et~al.} 2010, \apj,
  717, 183

\bibitem[{{Riess} {et~al.}(1998){Riess}, {Filippenko}, {Challis},
  {Clocchiatti}, {Diercks}, {Garnavich}, {Gilliland}, {Hogan}, {Jha},
  {Kirshner}, {Leibundgut}, {Phillips}, {Reiss}, {Schmidt}, {Schommer},
  {Smith}, {Spyromilio}, {Stubbs}, {Suntzeff}, \& {Tonry}}]{RFC1998}
{Riess}, A.~G., {Filippenko}, A.~V., {Challis}, P., {et~al.} 1998, \aj, 116,
  1009

\bibitem[{{Schaefer} \& {Pagnotta}(2012)}]{SP2012}
{Schaefer}, B.~E. \& {Pagnotta}, A. 2012, \nat, 481, 164

\bibitem[{{Schmidt} {et~al.}(1998){Schmidt}, {Suntzeff}, {Phillips},
  {Schommer}, {Clocchiatti}, {Kirshner}, {Garnavich}, {Challis}, {Leibundgut},
  {Spyromilio}, {Riess}, {Filippenko}, {Hamuy}, {Smith}, {Hogan}, {Stubbs},
  {Diercks}, {Reiss}, {Gilliland}, {Tonry}, {Maza}, {Dressler}, {Walsh}, \&
  {Ciardullo}}]{SSP1998}
{Schmidt}, B.~P., {Suntzeff}, N.~B., {Phillips}, M.~M., {et~al.} 1998, \apj,
  507, 46

\bibitem[{{Shen} \& {Bildsten}(2009)}]{SB2009}
{Shen}, K.~J. \& {Bildsten}, L. 2009, \apj, 699, 1365

\bibitem[{{Shen} {et~al.}(2010){Shen}, {Kasen}, {Weinberg}, {Bildsten}, \&
  {Scannapieco}}]{SKW2010}
{Shen}, K.~J., {Kasen}, D., {Weinberg}, N.~N., {Bildsten}, L., \&
  {Scannapieco}, E. 2010, \apj, 715, 767

\bibitem[{{Shen} \& {Moore}(2014)}]{SM2014}
{Shen}, K.~J. \& {Moore}, K. 2014, \apj, 797, 46

\bibitem[{{Sim} {et~al.}(2010){Sim}, {R{\"o}pke}, {Hillebrandt}, {Kromer},
  {Pakmor}, {Fink}, {Ruiter}, \& {Seitenzahl}}]{SRH2010}
{Sim}, S.~A., {R{\"o}pke}, F.~K., {Hillebrandt}, W., {et~al.} 2010, \apjl, 714,
  L52

\bibitem[{{Spruit}(2002)}]{S2002}
{Spruit}, H.~C. 2002, \aap, 381, 923

\bibitem[{{Staff} {et~al.}(2012){Staff}, {Menon}, {Herwig}, {Even}, {Fryer},
  {Motl}, {Geballe}, {Pignatari}, {Clayton}, \& {Tohline}}]{SMH2012}
{Staff}, J.~E., {Menon}, A., {Herwig}, F., {et~al.} 2012, \apj, 757, 76

\bibitem[{{Suh} {et~al.}(2011){Suh}, {Yoon}, {Jeong}, \& {Yi}}]{SYJY2011}
{Suh}, H., {Yoon}, S.-c., {Jeong}, H., \& {Yi}, S.~K. 2011, \apj, 730, 110

\bibitem[{{Suijs} {et~al.}(2008){Suijs}, {Langer}, {Poelarends}, {Yoon},
  {Heger}, \& {Herwig}}]{SLP2008}
{Suijs}, M. P.~L., {Langer}, N., {Poelarends}, A.-J., {et~al.} 2008, A\&A, 481,
  L87

\bibitem[{{Taam}(1980{\natexlab{a}})}]{T1980}
{Taam}, R.~E. 1980{\natexlab{a}}, \apj, 237, 142

\bibitem[{{Taam}(1980{\natexlab{b}})}]{T1980b}
{Taam}, R.~E. 1980{\natexlab{b}}, \apj, 242, 749

\bibitem[{{Tayler}(1973)}]{T1973}
{Tayler}, R.~J. 1973, \mnras, 161, 365

\bibitem[{{Theuns} {et~al.}(1996){Theuns}, {Boffin}, \& {Jorissen}}]{TBA1996}
{Theuns}, T., {Boffin}, H.~M.~J., \& {Jorissen}, A. 1996, \mnras, 280, 1264

\bibitem[{{van Kerkwijk} {et~al.}(2010){van Kerkwijk}, {Chang}, \&
  {Justham}}]{KCJ2010}
{van Kerkwijk}, M.~H., {Chang}, P., \& {Justham}, S. 2010, \apjl, 722, L157

\bibitem[{Waldman {et~al.}(2011)Waldman, Sauer, Livne, Perets, Glasner,
  Mazzali, Truran, \& Gal-Yam}]{WSL2011}
Waldman, R., Sauer, D., Livne, E., {et~al.} 2011, The Astrophysical Journal,
  738, 21

\bibitem[{{Wang} {et~al.}(2013){Wang}, {Justham}, \& {Han}}]{WJH2013}
{Wang}, B., {Justham}, S., \& {Han}, Z. 2013, A\&A, 559, A94

\bibitem[{{Warner}(1995)}]{W1995}
{Warner}, B. 1995, \apss, 225, 249

\bibitem[{{Woosley} \& {Kasen}(2011)}]{WK2011}
{Woosley}, S.~E. \& {Kasen}, D. 2011, \apj, 734, 38

\bibitem[{{Woosley} {et~al.}(1986){Woosley}, {Taam}, \& {Weaver}}]{WTW1986}
{Woosley}, S.~E., {Taam}, R.~E., \& {Weaver}, T.~A. 1986, \apj, 301, 601

\bibitem[{{Woosley} \& {Weaver}(1994)}]{WW1994}
{Woosley}, S.~E. \& {Weaver}, T.~A. 1994, \apj, 423, 371

\bibitem[{{Yoon} \& {Langer}(2003)}]{YL2003}
{Yoon}, S.-C. \& {Langer}, N. 2003, \aap, 412, L53

\bibitem[{{Yoon} \& {Langer}(2004{\natexlab{a}})}]{YL2004b}
{Yoon}, S.-C. \& {Langer}, N. 2004{\natexlab{a}}, \aap, 419, 645

\bibitem[{{Yoon} \& {Langer}(2004{\natexlab{b}})}]{YL2004a}
{Yoon}, S.-C. \& {Langer}, N. 2004{\natexlab{b}}, \aap, 419, 623

\bibitem[{{Yoon} \& {Langer}(2005)}]{YL2005}
{Yoon}, S.-C. \& {Langer}, N. 2005, \aap, 435, 967

\bibitem[{{Yuan} {et~al.}(2013){Yuan}, {Kobayashi}, {Schmidt}, {Podsiadlowski},
  {Sim}, \& {Scalzo}}]{YKS2013}
{Yuan}, F., {Kobayashi}, C., {Schmidt}, B.~P., {et~al.} 2013, \mnras, 432, 1680

\bibitem[{{Zapartas} {et~al.}(2017){Zapartas}, {de Mink}, {Izzard}, {Yoon},
  {Badenes}, {Gotberg}, {de Koter}, {Neijssel}, {Renzo}, {Schootemeijer}, \&
  {Shrotriya}}]{ZMI2017}
{Zapartas}, E., {de Mink}, S.~E., {Izzard}, R.~G., {et~al.} 2017, ArXiv
  e-prints

\end{thebibliography}
\end{document}